\documentclass[11pt]{article}
  
\usepackage{amsmath,amssymb}

\usepackage{epsfig}  
\usepackage{graphicx}               
\usepackage{url}
\usepackage{cite}
\usepackage{hyperref}
\usepackage{bm}
\usepackage{cancel}
\usepackage{pstricks}
\usepackage{color}
\usepackage{braket}

\newcommand{\nc}{\newcommand}  

\nc{\beq}{\begin{equation}}  
\nc{\eeq}{\end{equation}}  
\nc{\beqa}{\begin{eqnarray}}  
\nc{\eeqa}{\end{eqnarray}}  
\nc{\bit}{\begin{itemize}}  
\nc{\eit}{\end{itemize}}  

\DeclareMathOperator{\Tr}{Tr}

\DeclareMathOperator{\GeV}{GeV}
\DeclareMathOperator{\MeV}{MeV}
\DeclareMathOperator{\TeV}{TeV}

\newcommand\Xtilde{\stackrel{\sim}{\smash{X}\rule{0pt}{1.0ex}}}
\newcommand\Xbtilde{\stackrel{\sim}{\smash{\overline{X}}\rule{0pt}{1.4ex}}}
\newcommand\Phibtilde{\stackrel{\sim}{\smash{\overline{\Phi}}\rule{0pt}{1.4ex}}}

\title{
\vspace*{-2.3cm}
\begin{flushright}
\normalsize{
  }
\end{flushright}
\vspace{1.5cm}
\Large  
\textbf{Supersymmetric Resonant Dark Matter: \vspace*{0.1cm} \\ {\large a Thermal Model for the AMS-02 Positron Excess}} \vspace*{1.0cm}   
}

\author{Yang Bai, Joshua Berger and Sida Lu
\vspace{5mm}
\\
\normalsize\emph{Department of Physics, University of Wisconsin-Madison, Madison, WI 53706, USA}  
}
\date{}

\begin{document}  
\setcounter{page}{0}  
\maketitle  

\vspace*{1cm}  
\begin{abstract} 
We construct a thermal dark matter model with annihilation mediated by a resonance to explain the positron excess observed by PAMELA, Fermi-LAT and AMS-02, while satisfying constraints from cosmic microwave background (CMB) measurements. The challenging requirement is that the resonance has twice the dark matter mass to one part in a million. We achieve this by introducing an $SU(3)_f$ dark flavor symmetry that is spontaneously broken to $SU(2)_f \times U(1)_f$. The resonance is the heaviest state in the dark matter flavor multiplet and the required mass relation is protected by the vacuum structure and supersymmetry from radiative corrections. The pseudo-Nambu Goldstone Bosons (PNGB's) from the dark flavor symmetry breaking can be slightly lighter than one GeV and dominantly decay into two muons just from kinematics, with subsequent decay into positrons. The PNGB's are produced in resonant dark matter semi-annihilation, where two dark matter particles annihilate into an anti-dark matter particle and a PNGB. The dark matter mass in our model is constrained to be below around 1.9 TeV from fitting thermal relic abundance, AMS-02 data and CMB constraints. The superpartners of Standard Model (SM) particles can cascade decay into a light PNGB along with SM particles, yielding a correlated signal of this model at colliders. One of the interesting signatures is a resonance of a SM Higgs boson plus two collimated muons, which has superb discovery potential at LHC Run 2. 
\end{abstract} 
  
\thispagestyle{empty}  
\newpage  
  
\setcounter{page}{1}

\vspace{-2cm}

\section{Introduction}
\label{sec:intro}
It is beyond doubt that the majority of matter in the Universe is composed of dark matter, yet we still don't know how to describe the particle properties, {\it if any}, of dark matter as we can with other particles in the Standard Model (SM). The mechanism by which the abundance of observed dark matter is generated is not known either, though thermal freeze-out has long been regarded as the simplest explanation of the dark matter relic abundance. For order one coupling strength between dark matter particles and SM particles or other mediators, the dark matter mass is anticipated to be around the TeV scale in freeze-out models. These models furthermore generically predict additional contributions to the cosmic ray spectra of electrons/positrons, protons/anti-protons, photons and neutrinos, generated by dark matter annihilations in the present day.  Among the experimental searches for such cosmic rays, known as indirect detection searches, AMS-02 has provided the most precise measurement of the electron and positron energy spectrum up to 1 TeV~\cite{Aguilar:2014fea}. Their measurement of the positron fraction shows an excess above the standard background estimation up to an energy of 0.6 TeV~\cite{Accardo:2014lma,Ting:Dec2016Talk}. This interesting excess has also been seen in earlier experiments including HEAT~\cite{Beatty:2004cy}, PAMELA~\cite{Adriani:2013uda}, Fermi-LAT~\cite{FermiLAT:2011ab}. 

On the one hand, the energy scale of the positron excess matches the generic mass scale of thermal dark matter models, which provides a strong hint that the positron excess may be explained by thermal dark matter annihilations. On the other hand, the preferred annihilation rate from data, ${\cal O}(10^{-23}\mbox{cm}^3/\mbox{s})$, is two to three orders of magnitude higher than the required rate, $\sim 3 \times 10^{-26}\mbox{cm}^3/\mbox{s}$, of the simple $s$-wave annihilation thermal dark matter models. Additional complications are required in the dark matter sector to explain why the dark matter annihilation rate is higher in the present day Milky Way halo than during the time of thermal freeze-out. One frequent approach to accommodate both annihilation rates utilizes the so-called ``Sommerfeld enhancement'', whereby attractive long-range interactions among the dark matter particles yield a $1/v$ enhancement to the cross section for short-range annihilation. Because the dark matter averaged velocity in Milky Way is around $10^{-3}$, two orders of magnitude smaller than that during thermal freeze-out, the large present day annihilation rate preferred by data can be naturally explained~\cite{ArkaniHamed:2008qn,Pospelov:2008jd}. For this class of models, dark matter annihilation rates during the recombination era, where $v \ll 10^{-3}$, are further enhanced and dump energetic electrons and positrons into the plasma, which interact with the CMB photons and lead to excluded distortions of the CMB power spectrum (see Refs.~\cite{Cirelli:2016rnw,Bringmann:2016din} recent analysis for this class of models and Refs.~\cite{Chen:2003gz,Padmanabhan:2005es,2012PhRvD..85d3522F,Slatyer:2015jla} and Planck constraints~\cite{Ade:2015xua} for general models). Up to the model-dependent absorption efficiencies of electrons and positrons energy, the constraint on the dark matter annihilation rate is $\langle \sigma v \rangle \lesssim{\cal O}(10^{-24}\mbox{cm}^3/\mbox{s})$. 

\begin{figure}[th!]
\begin{center}
\includegraphics[width=0.6\textwidth]{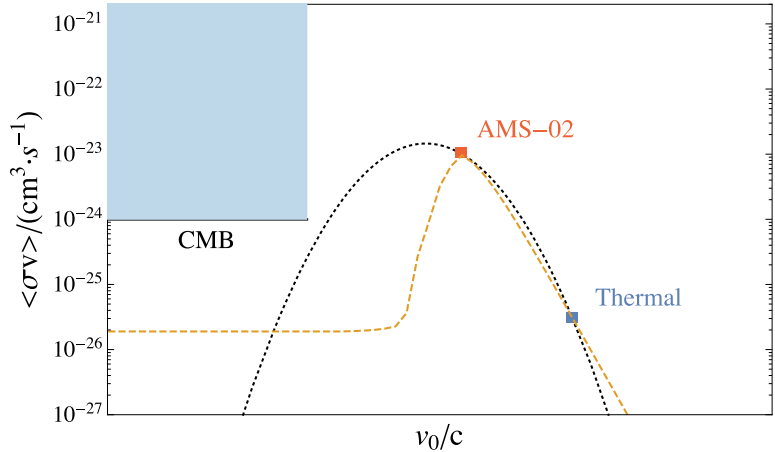}
\caption{A schematic plot to show the required annihilation rates for AMS-02 positron fraction data and dark matter thermal relic abundance, as well as the constraints of the CMB power spectrum from Planck. The dashed orange line is the possible behavior from resonance-mediated annihilation, while the dotted black line is a phenomenological fit to the data.}
\label{fig:schematic}
\end{center}
\end{figure}

Before we move to discuss other possible models, we first study the schematic picture told by the experimental data in Fig.~\ref{fig:schematic}. 
Because of the relations among three characteristic velocities, $v_{\rm CMB} < v_{\rm AMS} < v_{\rm thermal}$ and the large rate required for the AMS-02 data, the underlying dark matter annihilation rate $\langle \sigma v \rangle$ has to have a peak structure around $v_{\rm AMS}$. This observation is based on a simplified early universe model for the dark matter relic abundance. More possibilities are allowed within some non-standard cosmological models. From the particle physics point of view, the simplest explanation for the peak in Fig.~\ref{fig:schematic} is to have the dark matter annihilation mediated by a resonance. If the resonance particle mass is very close to twice of the dark matter mass, then the dark matter may reach its peak annihilation rate around $v_{\rm AMS}$, yielding a much larger rate for the dark matter annihilation in the Milky Way halo. 

Several studies of so-called ``Breit-Wigner Enhancement''~\cite{Ibe:2008ye,Guo:2009aj,Ibe:2009dx,Bi:2009uj,Bi:2011qm,An:2012uu,Duch:2017nbe} phenomenological models exist.  Such models can simultaneously fit to both the AMS-02 data and thermal relic abundance. For some models considered in the literature (for instance Ref.~\cite{Ibe:2009dx}), a light PNGB exists with a mass below one GeV and mainly decay into two electrons and two muons, which can satisfy gamma ray experimental constraints~\cite{Boudaud:2014dta,Lopez:2015uma,Scaffidi:2016ind} from Fermi-LAT~\cite{Ackermann:2013yva}. Although this class of models seem to be in the right direction to provide a thermal dark matter explanation for the data, it suffers its own problem from the theoretical point of view. The biggest issue with resonant annihilation is explaining why the resonance mass $M_{\rm R}$ is close to twice of the dark matter mass $M_X$ at per million level: $(M_{\rm R} - 2 M_X)/(2 M_X)  \sim {\cal O}(v_{\rm AMS}^2) \sim {\cal O}(10^{-6})$.

One of the simplest ways to explain the small mass splitting is to have the resonance be a bound state of two dark matter particles. This scenario requires a long-range force to provide the binding energy and also suffers the additional $1/v$ Sommerfeld enhancement for the annihilation rate at the CMB era. To solve this problem, one could have two interacting dark matter states, $X_1$ and $X_2$, with the mass splitting $\delta M \sim 1$~MeV and only long-range force for the $X_2$ state. Then, the lighter dark matter states $X_1$ can annihilate via exchanging an $X_2$ bound state in the $s$-channel. To have a natural model without fine-tuning the parameter space, the gauge coupling, if mediated by $U(1)'$ force, has to have the binding energy ${\cal O}(\alpha'^{2} M_X)$ match to the dark matter kinetic energy ${\cal O}(1~\mbox{MeV})$, so $\alpha' \sim 10^{-3}$. However, the additional gauge interaction for the $X_2$ can also shift its mass at loop level, with a contribution ${\cal O}(\alpha' M_X) \sim {\cal O}(1~\mbox{GeV})$, dramatically higher than the required small mass splitting ${\cal O}(1~\mbox{MeV})$. Another option is a higher-dimensional model such as the Universal Extra Dimension (UED) model~\cite{Appelquist:2000nn} (as mentioned in Ref.~\cite{Ibe:2008ye}). The second Kaluza-Klein (KK) mode has the mass close to twice the first KK-mode mass. This seems to be an interesting way to construct a UV model for resonant dark matter. However, additional quantum corrections are anticipated to generate at least loop suppressed brane-localized kinetic terms~\cite{Cheng:2002iz}. Taking those terms into account, if the couplings in the dark sector are large, then the ratio of the second KK-mode mass over the first KK-mode mass can be close to two only at $10^{-3}$ level and still far from phenomenologically needed $10^{-6}$ level. Small couplings may be possible from a low energy phenomenology perspective, but the implementation of boundary conditions could require some large couplings that would reintroduce problematically large corrections to the mass spectrum. It becomes a non-trivial model-building challenge to obtain a natural model to realize resonant dark matter annihilation. 

In this paper, we explore a new and natural way to realize resonant dark matter annihilation based on the symmetry breaking vacuum structure of non-Abelian global symmetry. As worked out a long time ago, the renormalizable potential of a certain representation of $SU(N)$ symmetry has only a few discrete vacuum structures~\cite{Li:1973mq}. For instance, one could have $SU(3) \rightarrow SU(2)\times U(1)$ or $SU(3) \rightarrow SU(2)$, depending on the coupling relations in the potential. Those symmetry breaking pattens are fairly stable against radiative correction and higher-order dimensional operator correction. The breaking pattern $SU(3) \rightarrow SU(2)\times U(1)$ is achieved, for instance, when an $SU(3)$ octet scalar gets a vacuum expectation value (VEV). The order parameter is $\langle \Phi \rangle \propto \mbox{diag}(1, 1, -2)$ after a particular choice of $SU(3)$ basis. If this order parameter spurion couples to dark matter fields in $3(\overline{3})$ representation, the ratio of the masses of the heavier dark matter state over the lighter dark matter states is therefore two. In order to have resonant annihilation, at least one of the dark sector states should be a boson. Spin one dark matter would have additional model building difficulties, so we are left with scalar dark matter at the TeV scale.  Additional symmetries are needed to explain why such a scalar is light, as it has a hierarchy problem analogous to that of the SM Higgs boson. Supersymmetry (SUSY) remains a leading candidate to solve the hierarchy problem, so we construct a ``Supersymmetric Resonant Dark Matter'' (SRDM) model, based on the dark matter flavor symmetry breaking of $SU(3)_f/SU(2)_f\times U(1)_f$.

One interesting coincidence is that the symmetry breaking of $SU(3)_f/SU(2)_f\times U(1)_f$ also provides a PNGB supermultiplet. The PNGB states could naturally have a mass at scale dramatically below the dark matter mass scale. If their mass is below around 1 GeV, the leading decays into SM particles will likely be two muons or two electrons, just from kinematics, which are functionally the ``best'' dark matter annihilation channels~\cite{Boudaud:2014dta,Lopez:2015uma,Scaffidi:2016ind}. The SRDM model thus solves two problems at once, providing further motivation for its structure.

The remainder of the paper is organized as follows. In Section~\ref{sec:intro}, we explicitly write down the necessary superpotential and soft terms for the dark matter states and the interactions to break the global symmetry. We then calculate the particle spectrum and decay rates in Section~\ref{sec:spectra}. In Section~\ref{sec:annihilations}, we calculate the annihilation cross section for the processes mediated by the resonance and the corresponding dark matter relic abundance. We show the parameter space to fit the AMS-02 and CMB data in Section~\ref{sec:ams-02} and show additional signals of the SRDM model in Section~\ref{sec:add-signals}. We conclude our paper in Section~\ref{sec:conclusion}.

\section{The Model based on $SU(3)_f/SU(2)_f\times U(1)_f$ Symmetry Breaking}
\label{sec:intro}
As is well known for non-supersymmetric theories, certain renormalizable potential of some representations of global Lie group can only have some finite possible vacuum symmetries. For $SU(3)_f$ with an adjoint representation $\Phi\equiv \Phi_a t_a$ with $t_a$ as the generators and $a = 1, \cdots, 8$, one possible vacuum is  $\langle \Phi \rangle \propto \mbox{diag}(1, 1, -2)$ with the unbroken symmetry $SU(2)_f\times U(1)_f$. If the unbroken symmetry is nearly exact in the low energy theory, the vacuum structure of $\Phi$ should stay fairly stable and is not easily modified by quantum correction or higher-dimensional operators. If this symmetry-breaking spurion, $\Phi$, couples linearly to other matter fields that is fundamental under $SU(3)_f$, the $SU(2)_f$-doublet field should have a mass half of the $SU(2)_f$-singlet field, which is exactly the required condition to realize resonant dark matter or Breit-Wigner features for dark matter annihilations. In this section, we build the SRDM model based on this simple observation in a supersymmetric theory. The reason that we choose a supersymmetric model is to extend the factor of two mass relation to particles with different spins or different parities. 

In the SRDM model, we have the spontaneously breaking global dark matter flavor symmetry to be $SU(3)_f$, under which we have two superfields, $X$ and $\overline{X}$, as 3 and $\overline{3}$ and one superfield $\Phi$ as 8. Based on the global symmetry, we have following renormalization interactions in the superpotential
\beqa
W_{SU(3)}&=&- y\,\overline{X}_{i}\Phi^i_j X^{j} \,+\, \mu_\Phi \Tr (\Phi \Phi) + \lambda_\Phi \Tr(\Phi \Phi \Phi) \,.
\label{eq:superpotential}
\eeqa
where $i=1,2,3$ are the indices of $SU(3)_f$. For simplicity, we assume that all parameters are positive and this superpotential conserves $P$ and $CP$. Minimizing the potential, one can have two degenerate supersymmetric minima for $\langle X \rangle = \langle \overline{X} \rangle=0$ with $\langle \Phi \rangle =0$ or $\langle \Phi \rangle = 4 \mu_\Phi/(\sqrt{3}\lambda_\Phi) t^8$, up to an arbitrary $SU(3)_f$ transformation. In the symmetry breaking vacuum,  $SU(3)_f$ breaks to its subgroup $SU(2)_f\times U(1)_f$ with two massless complex superfields that are doublets under $SU(2)_f$. If there is no additional interactions in the superpotential or soft potential, one has simple relations among the three dark matter states: $M_{X_3} = 2 M_{X_1} = 2 M_{X_2}$, which is simply from the symmetry breaking pattern of dark flavor symmetry.  Note that we have neglected a few terms, such as $\overline{X} X$.  The additional terms could be forbidden by a ${\cal Z}_2$ under which $X$ and $\Phi$ are charged, for example, which $\lambda_\Phi$ breaks by a small amount.  Furthermore, since this interaction would be in the superpotential, it is not radiatively generated in the absence of SUSY breaking effects.

Other than the $SU(3)_f$-conserving superpotential, we also need to introduce an explicit $SU(3)_f$ breaking superpotential for three purposes: (a) introducing interactions among the dark matter states; (b) providing masses for the pseudo-Nambu-Goldstone bosons and fermions (PNGB/F) of the coset space of $SU(3)_f/SU(2)_f \times U(1)_f$; (c) introducing interactions of the dark matter sectors with the SM sector. We introduce the following $SU(3)_f$ breaking potential with the three terms fulfilling the three goals above respectively
\beqa
W_{\cancel{SU(3)}} &\supset \frac{\lambda_X}{2}\,(X_1 X_1 X_3 + \overline{X}_1 \overline{X}_1 \overline{X}_3  ) +    \epsilon_1 \, \mu_\Phi^2\,  \Phi_8 \, + \, \epsilon_2 \,\sum_a \Phi^a H_u H_d  \,.
\label{eq:superpotential-breaking}
\eeqa
For the first term, one could also have interactions like $X_2 X_2 X_3$ or $X_1 X_2 X_3$. Since we will have the lightest stable dark matter state in $X_1$, those interactions are not important for dark matter phenomenology. We also note that there is a ${\cal Z}_3$ symmetry for the two superpotentials in Eqs.~(\ref{eq:superpotential}) and (\ref{eq:superpotential-breaking}), under which $X_i \rightarrow \omega X_i$, $\overline{X}_i \rightarrow \omega^2 \overline{X}_i$ and $\Phi \rightarrow \Phi$ with $\omega=e^{i 2\pi/3}$. This ${\cal Z}_3$ symmetry is sufficient to protect the dark matter states from decaying in the SRDM model.  The second term in the above superpotential can provide masses for the PNGB/F. $SU(3)_f$ freedom allows us to choose a basis where this term goes as $\Phi^8$ and $\Phi^3$, though we neglect the small correction due to a $\Phi^3$ term in what follows. This pushes the vacuum into the $\Phi^8$ direction, whereas it was previously arbitrary up to an $SU(3)_f$ rotation. Here, the dimensionless parameter, $\epsilon_1 \ll 1$, will perturb the vacuum expectation value (VEV) of $\Phi$ and introduce tiny mass differences among different flavor components of $X(\overline{X})$. For the last term in Eq.~(\ref{eq:superpotential-breaking}), the coefficient $\epsilon_2 \ll 1$ is introduced to mediate interactions of the dark sector to the SM sector. This term has negligible effects on vacuum structures and spectra of the SRDM model and will be only responsible for the PNGB/F decays to SM particles.  

As in the MSSM sector, various soft-mass terms could exist for the dark matter sector. One necessary soft term is needed to break the degeneracy of two vacua with $\langle \Phi \rangle =0$ and $\langle \Phi \rangle \neq 0$. We choose this term to respect the $SU(3)_f$ global symmetry. The second soft term that we also need is to break the degeneracy of the two dark matter states $X_1$ and $X_2$ and potentially have $X_1$ as the lightest dark matter state. They are
\beqa
V_{\rm soft} \supset - b_\Phi \, \mu_\Phi^2  \Tr(\Phi^2) \, + h. c.\, + \,b_{X_i} \, \mu_\Phi^2\,(X_i X_i^\dagger + \overline{X}_i \overline{X}_i^\dagger)  \,.
\eeqa
The dimensionless parameters $0< b_\Phi, b_{X_i} \ll \epsilon_1 \ll 1$ and will be determined later from our fit to the AMS-02 positron signal. 

To the first order in both $b_\Phi$ and $\epsilon_1$ and minimizing the potential, we have $\langle X \rangle = \langle \overline{X} \rangle =0$ and 
\beqa
f= \langle \Phi_8\rangle= \dfrac{4\sqrt{3} + 4\sqrt{3}b_\Phi + 3\lambda_\Phi \epsilon_1 }{3 \lambda_\Phi}\, \mu_\Phi\,.
\eeqa
It's easy to check that $V|_{\Phi=0} - V|_{\Phi=f} = (16 b_\Phi \mu_\Phi^4)/(3\lambda^2_\Phi) > 0$, the symmetry-breaking vacuum with $f\neq 0$ is the global minimum of the potential. Knowing the vacuum structure of the SRDM model, we first work out the particle spectrum and properties, followed by the dark matter annihilation rate. 

\section{Particle Mass Spectra and Decays}
\label{sec:spectra}
For a small soft parameters, the mass spectrum in the SRDM model is nearly supersymmetric. The tiny mass splittings among different components could be crucial for the dark matter phenomenology and also provide a natural model for the dark matter annihilation mediated by a resonance. The soft mass parameter modifies the scalar particle masses, while the fermion masses only change by the SUSY-breaking shift in $f$, so we first discuss the fermion masses and then come back to scalar masses.  

\subsection{Fermion Mass Spectrum}
Under the remaining approximately global symmetry $SU(2)_f\times U(1)_f$, three components of $\Xtilde$($\Xbtilde$) each separate into a doublet and a singlet. In the basis we choose, the doublets can be written as $\Xtilde_{\rm D} = (\Xtilde_{1}, \Xtilde_{2})^T$ and $\Xbtilde_{\rm D}$. The two singlets are $\Xtilde_{3} \equiv \Xtilde_{\rm R}$ and $\Xbtilde_{3}\equiv\Xbtilde_{\rm R}$. The Weyl fermions $\Xtilde$ and $\Xbtilde$ can be combined to form Dirac fermions. For the fermion fields contained in the superfield $\Phi$, there is an $SU(2)_f$ triplet field $\widetilde{\Phi}_{\rm T} = (\widetilde{\Phi}^1, \widetilde{\Phi}^2, \widetilde{\Phi}^3)^T$, two doublets as $\widetilde{\Phi}_{\rm D}(\Phibtilde_{\rm D}) = (\widetilde{\Phi}^4 \pm i\widetilde{\Phi}^5, \widetilde{\Phi}^6 \pm i\widetilde{\Phi}^7)^T$ and a singlet $\widetilde{\Phi}^8$. The two Weyl fermion doublets combine to form Dirac fermions, while the triplet and singlet are Majorana fermions.  In Table~\ref{mass_spec_fermions}, we show the mass square of various fermions in leading orders of $\epsilon_1$ and $b_\Phi$, where one can see that only the PNGF masses are suppressed by the small coefficients $\epsilon_1$ and $b_\Phi$. All other fermion masses are anticipated at the scale of $f$ for Yukawa couplings of order of unit. It also easy to check that the singlet fermion $\Xtilde_{\rm R}$ has its mass to be twice of the doublet fermion $\Xtilde_{\rm 1,2}$, as one anticipates.  
\begin{table}[hb!]
\begin{center}
\renewcommand{\arraystretch}{2.3}
	\begin{tabular}{c|c|c|c|c|c}
		\hline\hline
Fermions &  $\Xtilde_{1,2}$, $\Xbtilde_{1,2}$   &  $\Xtilde_{\rm R}$, $\Xbtilde_{\rm R}$ &  $\widetilde{\Phi}_{\rm T}$ &  $\widetilde{\Phi}_{\rm D}(\Phibtilde_{\rm D})$   & $\widetilde{\Phi}_8$  \\ \hline 
Mass  &   $M_X + \delta m$   &  $2 M_X + 2 \delta m$ & $\dfrac{3 \sqrt{3}}{4} \lambda_\Phi\, f$   & $\dfrac{4\,b_\Phi + \sqrt{3}\,\epsilon_1 \lambda_\Phi}{4}\mu_\Phi$  &   $\dfrac{\sqrt{3}}{4} \lambda_\Phi\, f$ \\
		\hline\hline
	\end{tabular}
	\caption{The fermion mass spectrum in the SRDM model. Here, the parameter $\delta m \equiv \sqrt{3}\, b_\Phi \lambda_\Phi f/8 = b_\Phi \, \mu_\Phi/2$ and $M_X \equiv y\,f/(2\sqrt{3}) - \delta m$.  We neglect some leading corrections in $\epsilon_1$ and $b_\Phi$ to the $\tilde{\Phi}_T$ and $\tilde{\Phi}_8$ masses that are unimportant to the remaining discussion.} 
	\label{mass_spec_fermions}
\end{center}
\end{table}

\subsection{Scalar Mass Spectrum}
\label{sec:spectrum}
The scalar fields $X$ and $\overline{X}$ separate into different parity states. The mass eigenstates $X_s$ and $X_p$ as
\beqa
\label{trans}
X_s = \dfrac{1}{\sqrt{2}}(X + \overline{X}^\dagger),\quad X_p = \dfrac{1}{\sqrt{2}}(X - \overline{X}^\dagger) \,.
\eeqa
Under the unbroken ${\cal Z}_3$ symmetry, we simply have $X_s \rightarrow \omega X_s$ and $X_p \rightarrow \omega X_p$. The subscripts $s$ and $p$ here indicate ``scalar'' and ``pseudo-scalar'' respectively. There are both doublet and singlet scalars and pseudo-scalars. We show the mass spectrum in Table \ref{mass_spec_x} at the leading order in $b_\Phi$, $b_{X_i}$ and $\epsilon_1$. 
\begin{table}[hb!]
\begin{center}
\renewcommand{\arraystretch}{2.2}
	\begin{tabular}{c|c|c|c|c|c|c}
		\hline\hline
Scalars &  $X^1_p$ &   $X^2_p$ &  $X^1_s$ &  $X^2_s$    &  $X^{\rm R}_s$ & $X^{\rm R}_p$     \\ \hline 
Mass &     $M_X$ &   $M_X + \delta m'$  &   $M_X + 2\,\delta m$  &   $M_X + 2\,\delta m + \delta m'$    &  $2\,M_X + \delta m + \delta m''$  &  $2\,M_X + 3\,\delta m + \delta m''$  \\
		\hline\hline
	\end{tabular}
	\caption{Mass spectrum of $X$ and $\overline{X}$ scalars. Here, $X^{\rm R}_{s, p} \equiv X^3_{s,p}$; $\delta m' \equiv \frac{1}{2}b_{X_2} \mu^2_{\Phi}/M_X$ and $\delta m'' \equiv \frac{1}{4}b_{X_3} \mu^2_{\Phi}/M_X$ with $b_{X_1} =0$. } 
	\label{mass_spec_x}
\end{center}
\end{table}

From Tables~\ref{mass_spec_fermions} and \ref{mass_spec_x}, it is easy to see that the state  $X^1_p$ is the lightest dark matter state for all positive dimensionless couplings and is the dark matter candidate in our model. For our later phenomenological studies, we will have all parameters $\delta m,  \delta m', \delta m'' = {\cal O}(\mbox{MeV})$ to be around the dark matter kinetic energy. The two heaviest scalars, $X^{\rm R}_s$ and $X^{\rm R}_p$, are just lightly heavier than twice of dark matter mass. We will show that one of them will be the relevant $s$-channel state to mediate resonant annihilations of dark matter particles. 

The scalar $\Phi$ soft potential contains terms like $\Phi \Phi$ and $\Phi^\dagger \Phi^\dagger$, which split the real and imaginary parts of the original scalar fields in the supermultiplets. To leading order in $b_\Phi$ and $\epsilon_1$, we have the mass spectrum shown in Table~\ref{mass_spec_phi}. 
\begin{table}[htb!]
\begin{center}
\renewcommand{\arraystretch}{2.2}
	\begin{tabular}{c|c|c|c|c|c}
		\hline\hline
Scalars &  $\Phi^{\rm T}_s$   &  $\Phi^{\rm T}_p$ &  $\Phi^{\rm D}_{s, p} = \frac{1}{\sqrt{2}}(\Phi^{\rm D} \pm \overline{\Phi}^{\rm D^\dagger})$ &  $\Phi^8_s$   & $\Phi^8_p$    \\ \hline 
Mass &   $m_{\widetilde{\Phi}^{\rm T}} - \delta m$   &   $m_{\widetilde{\Phi}^{\rm T}} + \delta m$ &  $\sqrt{m_{\widetilde{\Phi}^{\rm D}}(m_{\widetilde{\Phi}^{\rm D}} \mp 2 \, \delta m) }$ &   $m_{\widetilde{\Phi}^{8}} + \delta m$ & $m_{\widetilde{\Phi}^{\rm 8}} - \delta m$   \\
		\hline\hline
	\end{tabular}
	\caption{Mass spectrum of $\Phi$ scalars.} 
	\label{mass_spec_phi}
\end{center}
\end{table}
The light scalar fields are $\Phi^{\rm D}_{s, p}$ with their masses suppressed by $b_\Phi$ and $\epsilon_1$ at  linear order. 

Before we end this section, we briefly discuss the mass scales in our model. The dark matter mass, $M_X$ will be chosen to be ${\cal O}(1~\mbox{TeV})$. The doublet $X_{1,2}(\overline{X}_{1,2})$ fermions and scalars all have masses around $M_X$. The singlet $X_{\rm R}(\overline{X}_{\rm R})$ fermions and scalars have masses around $2 M_X$. The fermions and scalars in $\Phi$, excluding the Goldstone doublet superfield, have masses of ${\cal O}(\lambda_\Phi/y\, M_X)$ and could be dramatically heavier than the dark matter and decouple from dark matter phenomenology. The PNGB masses will be ${\cal O}(500~\mbox{MeV})$ to select muons as the leading decay channel from kinematics. The mass splittings, related to $\delta m, \delta m', \delta m''$, among different $X(\overline{X})$ states will be ${\cal O}(1~\mbox{MeV})$. Altogether, the small dimensionless parameters in our model have magnitudes
\beqa
\delta m = \frac{1}{2}\,b_\Phi \, \mu_\Phi = \frac{3}{4}\,b_\Phi\, \frac{\lambda_\Phi}{y} M_X  \sim 1~\mbox{MeV}  &\Rightarrow& b_\Phi\,  \sim 10^{-6}\times \frac{y}{\lambda_\Phi} \,, \\
\delta m' (\delta m'') = \frac{1}{2}\,b_{X_2}(b_{X_3}) \, \mu_\Phi^2/M_X = \frac{9}{16}\,b_\Phi\, \frac{\lambda_\Phi^2}{y^2} M_X  \sim 1~\mbox{MeV}  &\Rightarrow& b_{X_2}(b_{X_3})\,  \sim 10^{-6}\times \frac{y^2}{\lambda_\Phi^2} \,, \\
M_{\Phi^{\rm D}_{s,p}} \approx m_{\widetilde{\Phi}_D}  \approx \frac{3\sqrt{3}}{8} \epsilon_1 \, \frac{\lambda_\Phi^2}{y} \, M_X \sim 0.5~\mbox{GeV}  &\Rightarrow& \epsilon_1 \sim 10^{-3} \times \frac{y}{\lambda_\Phi^2} \,,
\eeqa
which justifies our assumption of perturbative calculations so far. 

\subsection{Interactions and Heavier Particle Decay Widths}
\label{sec:widths}
We have demonstrated so far that the state $X^1_p$ is the lightest stable states in our model. In principle, the supersymmetric partner states of $X^1_p$ and other states in the superfield of $X^2$ could also be stable on a cosmological time scale. Similarly, additional interactions are required to mediate the decay of the other states of the PNGB supermultiplet aside from $\Phi^{\rm D}_s$. Because of the ${\cal O}(\mbox{MeV})$ mass splitting among some states and in order for not having too much kinematic suppression factors, we introduce the following ``neutrino portal'' to mediate the heavier state decays. The relevant higher-dimensional operators in the superpotential are
\beqa
W_{\rm decay} \supset  \dfrac{1}{\Lambda_X}[a_{11} X_1 \overline{X}_1 + a_{12} X_1 \overline{X}_2 + a_{22} X_2 \overline{X}_2]\,L\,H_u + \dfrac{1}{\Lambda_\Phi}\Phi^{\rm D}\,\overline{\Phi}^{\rm D}\, L\,H_u \,,
\label{eq:higher-dim}
\eeqa
where we have ignored the flavor index for the SM leptons.  For the first term, we keep the different flavor couplings independent for the moment, for reasons that we discuss in detail below.

Based on the first operator in Eq.~(\ref{eq:higher-dim}), other states in the superfields $X^1$ and $X^2$ can either directly or in the cascade way decay into $X^1_p$. Assuming $\delta m^\prime > \delta m$, the allowed decay channels are $X^1_s, X^2_s, X^2_p \to \widetilde{X}^{1(2)}$ and $\widetilde{X}^{1(2)} \to X^1_p$. Their decay widths are calculated as
\beqa
\Gamma \left[X^i_{s(p)} \rightarrow \widetilde{X}^j + \nu_L\right] &=&\Gamma \left[X^i_{s(p)} \rightarrow \widetilde{X}^j + \overline{\nu_L} \right] = \dfrac{a_{ij}^2 v^2_u \left[M_{X^i_{s(p)}} - m_{\widetilde{X}^j}\right]^2}{16\pi M_X \Lambda^2_X} \,,\\
\Gamma(\widetilde{X}^i \rightarrow X^j_p + \nu_L) &=&\Gamma(\widetilde{X}^i \rightarrow X^j_p + \overline{\nu_L}) = \dfrac{a_{ij}^2 v^2_u\, (m_{\widetilde{X}^i} - M_{X^j_{p}})^2}{32\pi M_X \Lambda^2_X} \,,
\label{eq:X-d-decay}
\eeqa
where $i, j = 1, 2$, $v_u/\sqrt{2} = \braket{H^0_u}$. For a reasonably low cutoff scale, all heavier dark matter states decay fast to be treated as unstable particles for both dark matter thermal and indirect signal calculations. For instance, choosing $M_X = 1 \TeV$,  $\Lambda_X = 1000 \TeV$, $v_u \sim 246\GeV$, $a_{ij} = 1$ and a mass splitting of $1~\MeV$, the lifetime of these states are evaluated to be $\mathcal{O}(10^{-6}\,\mbox{s})$. We also note that both decays into neutrinos and anti-neutrinos can happen here because of the lepton-number violating interactions in our model.

For the real scalar of the PNGB, $\Phi^{\rm D}_s$, it's mixing with the $CP$-even Higgs in the MSSM can be derived from the two interaction terms, $\epsilon_2 \, \Phi^{\rm D} H_u H_d$ and $\mu H_u H_d$, in the superpotential. The mixing angle is at the order of $\epsilon_2 v_u \mu/M_h^2$. For the mass range of $2 m_\mu < M_{\Phi^{\rm D}_s} < 2 M_K$, the leading decay channel is 
\beqa
\label{eq:goldstone-decay}
\Gamma(\Phi^{\rm D}_s \rightarrow \mu + \overline{\mu} ) = \frac{\epsilon_2^2\, \mu^2\,m_\mu^2}{4 \pi \,  M_h^4} M_{\Phi^{\rm D}_s} \left(1 - \frac{4 m_\mu^2}{M_{\Phi^{\rm D}_s}^2}\right)^{3/2} \sim 1.8 \times 10^{-16}\,~\mbox{GeV} \,,
\eeqa
for $\mu \sim 1$~TeV, $\epsilon_2 \sim 10^{-4}$, and $M_{\Phi^{\rm D}_s} \sim 0.5$~GeV. The corresponding lifetime of this PNGB is around $10^{-11}\,\mbox{s}$. The superpartner states of $\Phi^{\rm D}_s$ can have decays mediated by the second operator in Eq.~(\ref{eq:higher-dim}) and have prompt decay widths for a not-too-high cutoff scale, $\Lambda_\Phi$. The main decay of the PNGF is $\widetilde{\Phi}^{\rm D} \rightarrow \Phi^{\rm D}_s + \nu_L(\overline{\nu_L})$ with its formula given by
\beqa
\Gamma[ \widetilde{\Phi}^{\rm D} \rightarrow \Phi_s^{\rm D} + \nu_L(\overline{\nu_L})] & = & \dfrac{ v^2_u (m_{\widetilde{\Phi}^{\rm D}} - M_{\Phi_s^{\rm D}})^2}{32\pi M_{\Phi_s^{\rm D}} \Lambda^2_\Phi} \,.
\eeqa
The decay of $\Phi_p^{\rm D}$ is $\Phi^{\rm D}_p \rightarrow \widetilde{\Phi}^{\rm D} + \nu_L(\overline{\nu_L})$ with its decay width calculated as
\beqa
\Gamma[\Phi_p^{\rm D} \rightarrow \widetilde{\Phi}^{\rm D} + \nu_L(\overline{\nu_L})] = \dfrac{ v^2_u (M_{\Phi_p^{\rm D}} - m_{\widetilde{\Phi}^{\rm D}})^2}{16\pi M_{\Phi_s^{\rm D}} \Lambda^2_\Phi} \,.
\label{eq:phi-d-p-decay}
\eeqa
Both of them have a lifetime of $10^{-9}$\,s for a mass splitting of order MeV, $M_{\Phi_s^{\rm D}}$~GeV and $\Lambda_\Phi \sim 1000$~TeV. 

Having discussed the decays of lighter dark matter states, we now turn to the properties of the heavy dark matter state, $X^3 \equiv X^{\rm R}$, which will play the role of the resonance in our model. The decays of the $X^{\rm R}$ singlet states can be related in the Breit-Wigner formalism to annihilation cross section of the $X^1_p$ dark matter state. From the kinematics and parity conservation, the only possible mass-on-shell state to mediate resonant annihilation of dark matter states is the $X^{\rm R}_s$ scalar, so we pay special attention to this particle. Its decay width into the two dark matter states is
\beqa
\Gamma(X_s^{\rm R} \rightarrow X^1_p + X^1_p) \approx \frac{\lambda_X^2\, M_X^2}{8 \pi \,\sqrt{s}}\,v_{\rm rel}\,,
\eeqa
with the $v_{\rm rel} = 2 \sqrt{1 - 4 M^2_X/s}$ ($s=M^2_{X_s^{\rm R}}$ for an on-shell resonance) or twice of the velocity of the particle in the final state. Another decay channel, $X_s^{\rm R} \rightarrow \widetilde{X}^{\rm D} + \widetilde{X}^{\rm D}$, is at higher order of $v_{\rm rel}$ and given by
\beqa
\Gamma(X_s^{\rm R} \rightarrow \widetilde{X}^{\rm D} + \widetilde{X}^{\rm D}) = \frac{1}{64\,\pi}\,\lambda_X^2\,\sqrt{s}\,v^{\prime 3}_{\rm rel} \,,
\eeqa
with $v^{\prime}_{\rm rel} = 2 \sqrt{1 - 4 M^2_{\widetilde{X}_D}/s}$ and suppressed. From an explicit calculation, one can show that there is no cubic interaction among $X^{\rm R}_s$ and two $X^1_s$ scalars. Finally, there are also decays into PNGB's and PNGF's, their decay branching widths are 
\beqa
\Gamma(X^{\rm R}_s \rightarrow X_s^1 + \Phi^{\rm D}_s ) = \Gamma(X^{\rm R}_s \rightarrow X_s^2 + \Phi^{\rm D}_s ) = \dfrac{3}{512\pi} y^2 M_X \,,\\
 \Gamma(X^{\rm R}_s \rightarrow X_p^1 + \Phi^{\rm D}_p )  = \Gamma(X^{\rm R}_s \rightarrow X_p^2 + \Phi^{\rm D}_p )  = \dfrac{27}{512\pi} y^2 M_X \,,
\\
\Gamma(X^{\rm R}_s \rightarrow \widetilde{X}^{1} + \widetilde{\Phi}^{\rm D})  = \Gamma(X^{\rm R}_s \rightarrow \widetilde{X}^{2} + \widetilde{\Phi}^{\rm D})  = \dfrac{9}{256\pi} y^2 M_X \,.
\eeqa
It is understood that only one component of the $\Phi^{\rm D}$ state enters the final state because of the $SU(2)_f$ symmetry for this decaying interaction.

\section{Resonant Annihilations}
\label{sec:annihilations}
In our model, the leading annihilation channels for dark matter at the current universe are dominated by the one mediated by an on-shell resonance. Specifically, we have the semi-annihilation processes of
\beqa
X^{\rm 1}_p + X^{\rm 1}_p \rightarrow X^{{\rm R}\dagger}_s \rightarrow   X_p^{{\rm 1,2} \dagger} + \Phi_p^{{\rm D} \dagger}, X_s^{{\rm 1,2} \dagger} + \Phi_s^{{\rm D} \dagger},  \Xbtilde_{\rm 1,2} + \Phibtilde_{\rm D} \,.
\eeqa
All the other states in $X^{\rm 1}$ and $X^2$ will decay into $X^1_p$. The states, $\Phi_p^{\rm D}$ and $\widetilde{\Phi}^{\rm D}$, decay to $\Phi_s^{\rm D}$, which decays to two muons in the SM. So, from the cosmic ray positron signal point of view, we can sum all the annihilation channels and group them together as the outgoing Breit-Wigner width. 

\begin{figure}[th!]
\begin{center}
\includegraphics[width=0.6\textwidth]{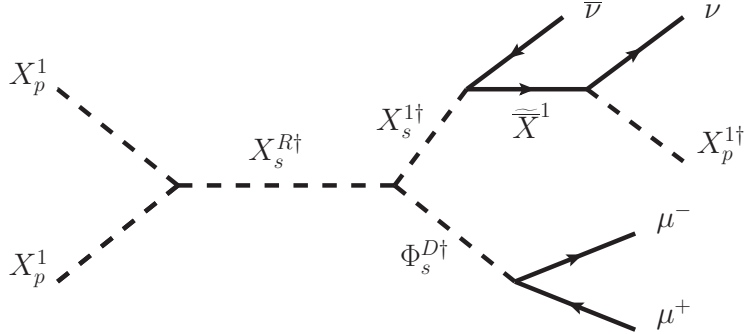}
\caption{A representative Feynman diagram for the semi-annihilation of dark matter mediated by a resonance. An unbroken ${\cal Z}_3$ is responsible for the stability of dark matter.  The muon energies can be up to the dark matter mass, while the neutrino energies are ${\cal O}(\mbox{MeV})$, determined by the mass difference of different dark matter states.}
\label{fig:feynman}
\end{center}
\end{figure}

Using the standard non-relativistic Breit-Wigner formula~\cite{Olive:2016xmw}, the dark matter annihilation cross section $X^{\rm 1}_p  X^{\rm 1}_p \rightarrow X_s^{{\rm R} \dagger} \rightarrow (X^{{\rm 1,2} \dagger}, \Xbtilde_{\rm 1,2})+(\Phi^{{\rm D}\dagger},  \Phibtilde_{\rm D})$ is given by
%
\beqa
\sigma_{X^1_p  X^1_p}(\sqrt{s}) = 2 \, \frac{(2 J + 1)}{(2 s_1 + 1)(2 s_2 + 1)} \, \frac{4\pi}{k\,k_{\rm in}} \left[ \frac{\Gamma^2/4}{(\sqrt{s}-M_{X^{\rm R}_s})^2 + \Gamma^2/4} \right]\,\mbox{B}_{\rm in}\, \mbox{B}_{\rm out} \,.
\label{eq:breit-wigner}
\eeqa
Here, the $J=0$ is the resonance spin; $s_1=s_2=0$ are the dark matter spin; $k = M_X v_{\rm rel}/2$ is the center-of-mass momentum in the initial state with $v_{\rm rel}$ as the relative speed of the two dark matter particles; $k_{\rm in} \approx M_X \sqrt{1 - 4 M^2_X/s}=M_X v_{\rm rel}/2$; the center-of-mass energy $s \approx 4 M_X^2 + M_X^2 v_{\rm rel}^2$; the resonance mass is $M_{X_s^{\rm R}} = 2 M_X + \delta m + \delta m''$.  We include an overall factor of two in Breit-Wigner formula to account for the fact that the initial state is made up of identical particles~\cite{törmä2014quantum}. The total width of the resonance as a function of $s$ is
\beqa
\Gamma(X_s^{\rm R}) = \frac{1}{16\pi}\,\lambda_X^2 \, M_X \, v_{\rm rel} \, + \, \frac{3}{16\,\pi} y^2\, M_X \,.
\eeqa
Introducing the resonance velocity, $v_{\rm R} \approx \sqrt{M_{X_s^{\rm R}}^2/M_X^2 - 4}\approx 2 \sqrt{(\delta m + \delta m'')/M_X}$, we have the effective annihilation rate
\beqa
(\sigma v_{\rm rel})_{\rm eff} = \frac{1}{2} \sigma_{X^1_p  X^1_p}(\sqrt{s})  = \frac{3\,y^2 \lambda_X^2}{4\pi\,M_X^2\,\left[ (v_{\rm rel}^2 - v_{\rm R}^2)^2 + \frac{1}{64\pi^2}\,(3\,y^2 + \lambda_X^2 v_{\rm rel})^2 \right]} \,,
\label{eq:breit-wigner-2}
\eeqa
where the factor of $1/2$ for the first equality comes from the fact that our dark matter field is a complex scalar instead of a real scalar.

In the narrow width approximation, the term inside the square bracket of Eq.~(\ref{eq:breit-wigner}) can be replaced by $\frac{\pi\,\Gamma}{2}\delta(\sqrt{s}-M_{X_s^{\rm R}})$. The annihilation rate takes the form of
\beqa
(\sigma v_{\rm rel})_{\rm eff} = 2\, \frac{8\pi}{M_X^2\, v_{\rm rel}} \, \frac{\pi \, \Gamma}{2}\, \mbox{B}_{\rm in}\, \mbox{B}_{\rm out} \, \delta(\sqrt{s}-M_{X_s^{\rm R}}) =  2\, \frac{8\pi}{M_X^2\, v_{\rm rel}} \, \frac{\pi \, \Gamma}{2}\, \mbox{B}_{\rm in}\, \mbox{B}_{\rm out}  \, \frac{2}{M_X\,v_{\rm rel}} \, \delta(v_{\rm rel} - v_{\rm R}) \,.
\label{eq:sigma-v-1}
\eeqa
Substituting the widths and branching ratios into Eq.~(\ref{eq:sigma-v-1}), we have the annihilation rate to be 
\beqa
(\sigma v_{\rm rel})_{\rm eff} = \frac{3\pi\,\lambda_X^2\, y^2}{v_{\rm rel}\,(3\, y^2 + \lambda_X^2\,v_{\rm rel})\,M_X^2} \delta(v_{\rm rel} - v_{\rm R}) \,.
\eeqa
%

\subsection{Annihilation Rates in our Galactic Halo and the CMB Era}
\label{sec:current-annihilation}

Assuming the normalized and isotropic dark matter velocity distribution is 
\beqa
f(v) = \frac{4\pi}{(\pi v_0^2)^{3/2}} \,e^{-v^2/v_0^2} \,,
\eeqa
such that $\int_0^\infty f(v) v^2 dv = 1$.  Here, $v_0 \approx 220$~km/s for the Standard
Halo Model~\cite{Vogelsberger:2008qb}. We ignore the distortions due to the escape velocity and will have the resonance velocity be close to $v_0$.  In terms of the two dark matter relative velocity, the normalized distribution is 
\beqa
f(v_{\rm rel}) = \frac{4\pi}{(2\pi v_0^2)^{3/2}} \,e^{-v_{\rm rel}^2/2v_0^2} \,.
\eeqa
Using the delta function approximation, the averaged dark matter annihilation rate is 
\beqa
\langle (\sigma v_{\rm rel})_{\rm eff} \rangle \equiv \int dv_{\rm rel}\, v_{\rm rel}^2\, (\sigma v_{\rm rel})_{\rm eff}\, f(v_{\rm rel}) =  \frac{3\pi\,\lambda_X^2\, y^2\,v_{\rm R}}{(3 y^2 + \lambda_X^2\,v_{\rm R})\,M_X^2} \, \frac{4\pi}{(2\pi v_0^2)^{3/2}} \,e^{-v_{\rm R}^2/2v_0^2} \,,
\eeqa
which provides a good approximation for the narrow-width case.

\begin{figure}[th!]
\begin{center}
\includegraphics[width=0.6\textwidth]{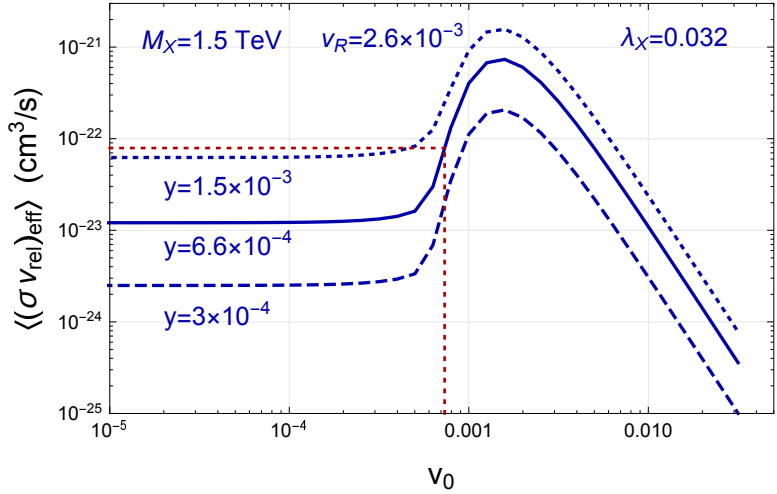}
\caption{Averaged annihilation rate as a function of the averaged dark matter velocity $v_0$. For the dark matter velocity in our galaxy halo with $v_0\approx 7.3\times 10^{-4}$, the averaged annihilation rate is $7.9\times 10^{-23}\,\mbox{cm}^3/\mbox{s}$ for the benchmark point with $y=6.6\times 10^{-4}$, while it is $1.2\times 10^{-23}\,\mbox{cm}^3/\mbox{s}$ for a small velocity at the CMB era.}
\label{fig:sigmav}
\end{center}
\end{figure}

In Fig.~\ref{fig:sigmav}, we show the averaged annihilation rate as a function of $v_0$ for fixed model parameters. For a benchmark model point with $M_X=1.5$~TeV, $v_{\rm R} = 2.6\times 10^{-3}$, $\lambda_X=0.032$ and $y=6.6\times 10^{-4}$, we have the prediction of $\langle\sigma v_{\rm rel}\rangle_{\rm eff}^{\rm AMS} = 7.9 \times 10^{-23}$~$\mbox{cm}^3/\mbox{s}$. For a small value of $v_0 \ll 10^{-4}$, relevant for the CMB era, the predicted annihilation rate is $\langle\sigma v_{\rm rel}\rangle_{\rm eff}^{\rm CMB} = 1.2 \times 10^{-23}$~$\mbox{cm}^3/\mbox{s}$, around a factor of seven smaller than the prediction for AMS-02. One can see that the resonant dark matter could provide a sufficiently large rate for AMS-02 and at the same time satisfy the constraints from CMB. To understand better of this good feature of resonance effect, we can set $v_{\rm rel}=0$ in Eq.~(\ref{eq:breit-wigner-2}) and have the averaged annihilation rate at the CMB era to be  
\beqa
\langle \sigma v_{\rm rel}\rangle_{\rm eff}^{\rm CMB} \approx \frac{3\,y^2\,\lambda_X^2}{4\,\pi\,M_X^2\,v_{\rm R}^4} \,,
\eeqa
and the ratio of the two annihilation rates as
\beqa
\frac{\langle \sigma v_{\rm rel}\rangle_{\rm eff}^{\rm CMB}}{\langle \sigma v_{\rm rel}\rangle_{\rm eff}^{\rm AMS}} = \frac{(3y^2 + \lambda_X^2 v_{\rm R})}{4\pi^2\,v_R^5}\,\frac{(2\pi v_0^2)^{3/2}}{4\pi} \,e^{v_{\rm R}^2/2v_0^2}  \,.
\eeqa
For $v_{\rm R} \sim v_0$, one can have a suppressed annihilation rate during the CMB era only when $y \lesssim v_{\rm R}$ and $\lambda_X \lesssim v_{\rm R}^{1/2}$. 

\subsection{Dark Matter Thermal Relic Abundance and Kinetic Decoupling}
\label{sec:thermal}
In order to determine the thermal relic abundance, we first note that resonant models are known to have the potential for early kinetic decoupling~\cite{Bi:2011qm,Duch:2017nbe}.  The completion of the process of thermal and kinetic freeze-out is slow, due to the increasing resonant enhancement of the semi-annihilation cross section.  For kinetic decoupling in our semi-annihilation case, this behavior is exaggerated by the large energy of the final state dark matter, which is produced with a momentum of order $3\,M_X/ 4$. If the semi-annihilation process dominates the evolution of the dark matter energy too early or if the dark matter fails to maintain kinetic equilibrium even with itself, then the dark matter could end up going through the resonance in semi-annihilation too late, leading to phenomenological issues such as an overproduction of dark matter or a large cross section around the time of recombination that is ruled out by CMB data. In order to satisfy these phenomenological constraints, we consider parameter space for the SRDM model that has a schematic cosmological evolution as follows:
\begin{itemize}
\item Dark matter loses chemical equilibrium and ceases to follow a Boltzmann suppressed number density.  The dark matter continues to deplete due to an enhanced semi-annihilation cross section approaching the resonance.  It maintains kinetic equilibrium dominantly via resonantly enhanced elastic scattering off neutrinos, effectively via the $a_{12}$ term in Eq.~(\ref{eq:higher-dim}).  The diagram for elastic scattering is shown in Fig.~\ref{fig:scatter}.
\begin{figure}[th!]
\begin{center}
\hspace*{-0.75cm}
\includegraphics[width=0.5\textwidth]{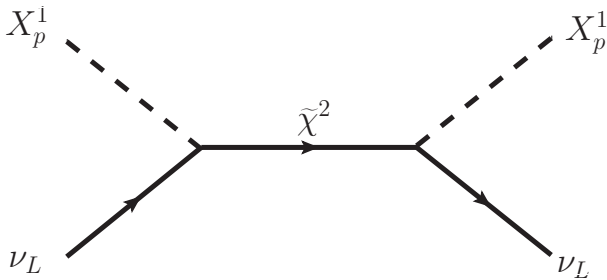} 
\caption{Feynman diagrams for the dark matter elastically scatter off neutrinos via a resonance in the $s$-channel.}
\label{fig:scatter}
\end{center}
\end{figure}
\item As the dark matter and SM bath temperature goes through the resonance at $T \sim 1~{\rm MeV}$ in both the semi-annihilation and neutrino scattering processes, these processes become significantly less efficient.  The abundance of dark matter per comoving volume freezes out to nearly its present day value.  There is a large initial energy dump into the dark matter at $x \gtrsim 1/\delta$, as kinetic equilibrium is lost and the semi-annihilation process comes to dominate the kinetic evolution of the dark sector. After this energy dump, the dark sector cools slowly as it redshifts. There are two possibilities for the dark matter evolution at this point. If a new mediator is introduced that couples only to the dark matter, then the dark matter may thermalize among itself. Otherwise, the quasi-relativistic dark matter produced in residual semi-annihilation comes to dominate and has only a small annihilation rate. The choice between the two possibilities has little relevance for indirect detection, CMB constraints or the relic abundance of dark matter, but may have other interesting cosmological consequences as the dark sector gains a large amount of extra energy from the residual semi-annihilation events.

\item As the dark matter slowly cools in either way, it eventually crosses through the semi-annihilation resonance pole once more, then decouples and cools rapidly compared to the SM sector, with only a small residual semi-annihilation rate.
\end{itemize}
In the remainder of this subsection, we study each of these pieces in turn.

The process of chemical decoupling is governed by the Boltzmann equation for the number density, $n$, of $X^1_p$.  It is convenient to define the quantity ``yield'', $Y \equiv n /s$ with the entropy $s = 2 \pi^2 g_* T^3/45$.  Then, the resulting Boltzmann equation for $Y$ due to the resonant semi-annihilation process~\cite{Griest:1990kh,DEramo:2010keq,Belanger:2012vp} is
\beq
\frac{dY}{dx} = - \frac{1}{2} \, \frac{s}{x \, H} \, Y \, \left(\langle \sigma \, v_{\rm rel} \rangle_{{\rm eff},x_{\rm DM}} Y - \langle \sigma \, v_{\rm rel} \rangle_{{\rm eff},x}  Y^{\rm eq}\right)  \,,
\label{eq:Y-equation}
\eeq
where for future convenience we define $x = M_X / T$ with $T$ as the SM particle temperature and $x_{\rm DM} = M_X / T_X$ for the dark matter temperature could be different from $x$ after the dark matter sector kinetically decouples from the SM sector. We include a factor of $1/2$ that arises due to having identical particles in the initial state phase space integration. The number density is the summation of dark matter and anti-dark matter number densities. In the radiation dominated era, $H = (8\pi \rho/3 M_{\rm pl}^2)^{1/2}$, $t = 1/(2 H)$, $\rho(T) = g_* \pi^2 T^4/30$ and $n^{\rm eq}(T) = g\,(M_X\, T/ 2\pi)^{3/2} e^{- M_X / T}$, where we will choose $g_* = 10.75$ because of the freeze-out temperature at a few MeV and $g=2$ for the dark matter degrees of freedom.  $Y^{\rm eq}\equiv n^{\rm eq} /s$ is the kinetic and chemical equilibrium of $Y$. The final dark matter relic abundance has $\Omega_{\rm DM} = M_X \, n/\rho_c = M_X\,s_0\,Y(\infty)/\rho_c$ with the critical density $\rho_c = 3 H_0 M_{\rm pl}^2/8\pi = 1.0539\times 10^{-5}\,h^2\,\mbox{GeV}\,\mbox{cm}^{-3}$ and $s_0 = 2889.2\,\mbox{cm}^{-3}$ as the entropy today. From the Planck collaboration, the measured dark matter energy density has $\Omega_{\rm DM} h^2 = 0.1199\pm0.0022$~\cite{Ade:2015xua}. 

For the dark matter thermal averaged annihilation rate and in terms of the parameter $x_{\rm DM}$, we have
\beqa
\langle\sigma v_{\rm vel}\rangle_{{\rm eff}, x_{\rm DM}} = \frac{x_{\rm DM}^{3/2}}{2\pi^{1/2} }\,\int^{\infty}_0 dv_{\rm rel} \,v_{\rm rel}^2 \, e^{- v_{\rm rel}^2\,x_{\rm DM}/4} \, (\sigma v_{\rm vel})_{\rm eff}  \,.
\eeqa
To understand the behavior during the different periods of the extended freeze-out process, we use the general parametrization for the resonance annihilation (see Appendix~\ref{sec:general-parametrization} for details), 
\beqa
(\sigma v_{\rm rel})_{\rm eff} = \sigma_0 \, \frac{\delta^2 + \gamma^2}{(\delta - v_{\rm rel}^2/4)^2 + \gamma^2} \,,
\label{eq:general-para}
\eeqa
with the resonance speed $v_{\rm R} = 2 \sqrt{\delta}$ and the parameter $\gamma$ related to the resonance width. During earlier times with $x \ll 1/\delta$, the resonance annihilation cross section is growing with $x_{\rm DM}$ and is given to very good approximation by
\beq
\langle \sigma \, v_{\rm rel} \rangle_{{\rm eff},x_{\rm DM}} = \sigma_0\,\frac{2 \sqrt{\pi}\,x_{\rm DM}^{3/2}\,\delta^{5/2}}{\gamma} \,.
\eeq
When the resonant semi-annihilation process is efficient, the dark matter follows its equilibrium distribution as usual, with $Y = Y^{\rm eq}$. The chemical equilibrium is lost when $\langle \sigma v_{\rm rel} \rangle_{\rm eff}\, n^{\rm eq} \lesssim H$ as in the usual $s$-wave or $p$-wave annihilation cases, and with the chemical decoupling $x_{\rm cd} \sim 20$ as usual.  

After chemical decoupling, since the cross section grows as the temperature approaches the resonant temperature at $x_{\rm DM} \rightarrow 1/\delta$, the dark matter number density continues to decrease as a power law, instead of rapidly freezing out to a plateau as in the standard freeze-out calculation.  At this point, $Y^{\rm eq} \ll Y$ and can be neglected, so that the differential equation in Eq.~(\ref{eq:Y-equation}) can be solved easily and provides a power-law behavior for $Y$ before the kinetic decoupling time $x_{\rm kd}$
\beq
Y \propto x^{-1/2}   \,, \qquad\qquad \mbox{for} \, \quad x_{\rm cd} < x < x_{\rm kd} \,.
\eeq
We have assumed at this point that kinetic coupling is maintained between dark matter and the SM sector. This is a non-trivial assumption and we consider this further now. At temperatures below the weak scale, there are two processes that could potentially maintain kinetic equilibrium: scattering off the relativistic $\Phi_{\rm D}$ states via quartic interactions as well as via $t$-channel $\Phi_3$ and $\Phi_8$ exchange or scattering off neutrinos via the first operator in Eq.~\eqref{eq:higher-dim}. The first of these possibilities necessarily is lost at $x \sim M_X / M_{\Phi_{\rm D}}$ as the $\Phi_{\rm D}$ become non-relativistic and rapidly decay away. We therefore consider parameter space where the second possibility is large. It in fact receives a resonant enhancement as well, as is clear from the structure of dominant diagram in Fig.~\ref{fig:scatter}. We consider only the dark matter flavor-changing superpotential operators $X_1 \overline{X}_2 L H_u$ to be significant. If the dark matter flavor-conserving operators have large coefficients, then they contribute to the neutrino Majorana mass at an unacceptably large level. Kinetic equilibrium is maintained until $x \sim 1/\delta$ for $\Lambda_X  / a_{12} \sim v_{\rm EW}$, the Higgs VEV. The operator should be UV completed at the TeV scale, though we do not study such a completion further in this work. To study the process of kinetic decoupling and freeze-out, we follow Refs.~\cite{Bringmann:2009vf,vandenAarssen:2012ag} and define a normalized measure of the dark matter temperature
\beq
y \equiv \frac{M_X\,T_{\rm DM}}{s^{2/3}} \equiv \frac{4\,g}{Y \, s^{5/3}} \int \frac{d^3p}{(2\,\pi)^3} \,\mathbf{p}^2 \,f_X(p) \,,
\eeq
where $f_X$ is the phase space distribution of dark matter.  The variable $y$ is defined such that after kinetic freeze-out, $y \to {\rm constant}$ and in kinetic equilibrium $y \propto x^{-1}$ up to changes in the composition of the SM relativistic fluid.  The evolution of $y$ is governed by a higher moment of the Boltzmann equation and is given in this era by~\cite{Bringmann:2009vf}
\beq
\frac{dy}{dx} = - \frac{1}{H\, x} \,2 \, M_X \, c(T) \, (y - y_{\rm EQ})  \,,
\label{eq:small-y-equation}
\eeq
with $c(T)$ defined and calculated in Appendix~\ref{sec:kinetic-functions}. We will choose the relevant model parameters in $c(T)$ to keep dark matter kinetically coupled to SM before its chemical freeze-out. 

At $x \gtrsim 1 / \delta$, the abundance of dark matter per comoving volume freezes out. At this point, kinetic equilibrium with the SM sector is lost as well. Since the abundance of dark matter is already frozen, this will turn out to not have a large effect on the phenomenology relevant to this work, provided that dark matter is sufficiently cold by the time the SM goes through recombination, which is the case regardless of the kinetic decoupling phenomenology. We therefore only briefly describe a couple of possibilities here and present some in more details in Appendix \ref{sec:kinetic-functions}. If there is insufficient interaction within the dark sector, then the dominant dark matter population will be produced with an energy very close to $3 M_X / 4$ as predicted from semi-annihilation with an initial state at rest. As the universe expands, this initial energy redshifts.  Even if there is sufficient interaction among dark matter particles, the energy dumped into the dark matter fluid due to residual semi-annihilations is significant. In either case, the dark sector does eventually cool back to a point where very few particles have a velocity above the resonant velocity, at which point the semi-annihilation process shuts off and the dark matter quickly redshifts to a very cold distribution.  At this point, the averaged dark matter semi-annihilation goes to its zero velocity level.  All of this happens before recombination in the SM sector, so it does not have any bearing on the CMB bound as derived above.

\begin{figure}[th!]
\begin{center}
\includegraphics[width=0.6\textwidth]{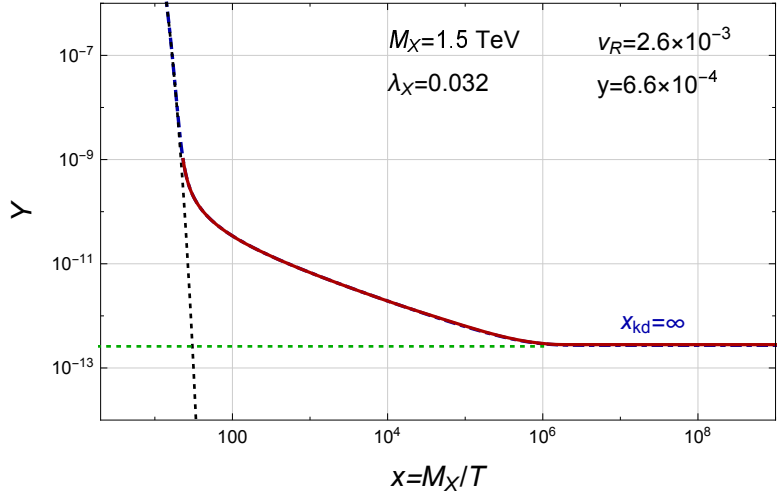}
\caption{The dark matter yield as a function of the temperature parameter $x$ for a benchmark point. The red line is the result from solving the full coupled Eqs.~(\ref{eq:Y-equation})(\ref{eq:small-y-equation})(\ref{eq:boltzmann-semi-ann}), while the blue line is using $x_{\rm DM} =x$ and only solving Eq.~(\ref{eq:Y-equation}). This benchmark point has $\Omega_{X} h^2 \approx 0.12$. }
\label{fig:yield}
\end{center}
\end{figure}

In Fig.~\ref{fig:yield}, we show the behavior of the yield function in terms of the temperature parameter $x$. We show the results from solving the full coupled differential equations of Eqs.~(\ref{eq:Y-equation})(\ref{eq:small-y-equation})(\ref{eq:boltzmann-semi-ann}) in the red curve and from the approximate approach in the blue curve. The good overlapping between those two curves justify our understanding that the kinetic decoupling in our model happens after the chemical freeze-out. Furthermore, the final freeze-out temperature is around 1~MeV and larger than the one without considering kinetic decoupling. We also note that the temperature of dark matter has a non-trivial evolution even after the chemical freeze-out. Since this is not important for the main purpose of our current paper, we show its behavior in Fig.~\ref{fig:xdm} of Appendix~\ref{sec:kinetic-functions}.

In Appendix~\ref{sec:general-parametrization}, we have used a general parametrization for the resonance annihilation in Eq.~(\ref{eq:general-para}) and found a simple relation of parameters to fit the thermal relic abundance
\beqa
\frac{\gamma}{\sigma_0 \,\delta^{2}}= 1.35 \times 10^{10}\, \mbox{GeV}^2 \,. 
\eeqa
Applying that to our model, the condition to satisfy the thermal dark matter relic abundance is approximately
\beqa
M_X^2\, \frac{3 y^2 + \lambda_X^2 \,v_R}{y^2\,\lambda_X^2}\,   \approx 2.0\times 10^{10}\,\mbox{GeV}^2 \,,
\label{eq:thermal-relation}
\eeqa
In later sections of our model, we will use this approximate relation to satisfy the dark matter thermal relic abundance for our parameter space. 

\section{Fit to AMS-02 Data and CMB}
\label{sec:ams-02}
The primary motivation for the models considered in this work are as explanation for the excess in the cosmic ray positron  flux above a few GeV, as seen at PAMELA~\cite{Adriani:2013uda}, Fermi-LAT~\cite{FermiLAT:2011ab}, and AMS-02~\cite{Accardo:2014lma}.  Dark matter annihilations have been considered as a potential explanation for the excess in the past.  Annihilations into quarks, gluons or electroweak gauge bosons can provide a good fit to the positron data alone, but there are additional stringent constraints from gamma ray searches at VERITAS~\cite{Aliu:2012ga}, Fermi-LAT~\cite{Ackermann:2013yva}, MAGIC~\cite{Aleksic:2013xea}, and H.E.S.S.~\cite{Abramowski:2014tra} that rule out these possibilities by orders of magnitude \cite{Boudaud:2014dta}.  Direct annihilations into electron pairs do not provide a good fit to the data \cite{Boudaud:2014dta}, while direct annihilations into $\mu$ and $\tau$ pairs provide good fits, but are also highly constrained by the same gamma ray searches.  The remaining possibility considered in the literature that can have a sufficiently large cross section, while providing a reasonable fit to the data, is annihilation into two light particles that each decay to a pair of leptons, generally referred to as $\phi\phi \to 4\ell$ models \cite{Boudaud:2014dta,Lopez:2015uma,Scaffidi:2016ind} with $\ell = e, \mu$.  Such models additionally have relatively weak constraints from annihilated products distorting the CMB spectrum~\cite{Slatyer:2009yq}, though the constraints are still significant and we revisit them below.  The models considered here effectively behave like $\phi\phi \to 4\ell$ benchmark models for sufficiently light PNGB's $\Phi^{\rm D}_s$ (see the annihilation Feynman diagram in Fig.~\ref{fig:feynman}), up to a remapping of the parameter space that we discuss below.  If the mass of the PNGB is above $2 m_K$, then significant hadronic modes open up generate conflicts with the data outlined above, so we consider $\Phi^{\rm D}_s$ masses below $2 m_K \sim 1~{\rm GeV}$ and the dominant decay into $\mu^+\mu^-$, which subsequently decay into $e^+e^-$ to increase the positron fraction. For our benchmark point, we will choose $M_{\Phi^{\rm D}_s}= 0.5$~GeV.

In order to fit the AMS-02 data, we calculate the contribution of our model to the positron and electron fluxes.  Noting that the energy of a $\Phi^{\rm D}_s$ particle produced in the annihilation channel studied in the model above is
\beq
\label{eq:phi-momentum}
E_{\Phi^{\rm D}_s} = \frac{3 M_X}{4} \,.
\eeq
from neglecting the tiny $\Phi^{\rm D}_s$ particle mass and the small dark matter kinetic energy. 
This is different from the ordinary annihilation case with $E_{\Phi^{\rm D}_s}= M_X$ in Refs.~\cite{Boudaud:2014dta,Lopez:2015uma,Scaffidi:2016ind}. We can then successively determine the spectrum of positrons from their parent chain as for example in Ref.~\cite{Elor:2015bho}.
The spectrum of positrons produced at the time and location of DM annihilation is given by
\beq
\label{eq:spectrum-prod}
\frac{dN}{dE} = \frac{4}{3\,M_X} \int_{\frac{4E}{3M_X}}^1 \,\frac{dx_\phi}{x_\phi} \, \int_{x_\phi}^1 \, \frac{dx_\mu}{x_\mu} \, \frac{dN}{dx_\mu} \,,
\eeq
where $x_\phi = 2 \, E_e^{\Phi^{\rm D}_{s}~{\rm rest}} / M_{\Phi_D}$, $x_\mu = 2 \, E_e^{\mu~{\rm rest}} / m_\mu$, and $dN / dx_\mu$ is the spectrum of positrons produced in a muon decay in the muon rest frame.  By $E_e^{P~{\rm rest}}$, we mean the energy of the positron in the rest frame of $P$.  This last spectrum can be calculated from the four-Fermi interaction for muon decay and is given by
\beq
\frac{dN}{dx_\mu} = 2\, x_\mu^2 \,(3 - 2\, x_\mu) \,,
\eeq
such that the resulting positron spectrum in the dark matter rest frame is given by
\beq
\frac{dN}{dE} = \frac{2}{27 \,M_X} (-8\, x^3 + 27\, x^2 - 30\, \log x - 19)  \,,
\eeq
with $x = 4E / 3M_X$ as the fraction of electron energy over its maximum energy.

The spectrum at annihilation is related to the spectrum observed at detectors at or near the Earth by propagation through the galactic medium.  For this work, we follow the simplified propagation model fit described in Ref.~\cite{Cirelli:2008id}.  In their formalism, the flux of positrons at Earth is given by
\beq
\frac{d\Phi}{dE} = \frac{v_e}{4\,\pi \, b(x)} \, \frac{1}{2} \, \left(\frac{\rho_{\odot}}{M_X}\right)^2 \, \int_E^{3M_X/4} dE^\prime \, \langle (\sigma v)_{\rm eff} \rangle \, \frac{dN}{dE^\prime} \, I\left[\lambda_D\left(E,E^\prime\right)\right],
\eeq
where $v_e \approx c$  is the electron velocity, $\rho_{\odot}=0.3~\mbox{GeV}/\mbox{cm}^3$ is the local dark matter energy density, and the energy loss function $b$, halo function $I$ and diffusion length $\lambda_D$ are discussed below.  Following Ref.~\cite{Cirelli:2008id}, we write the energy loss function as
\beq
b(E) = \frac{E^2}{{\rm GeV}\, \tau_E}  \,,
\eeq
with $\tau_E = 10^{16}~{\rm sec}$.  The diffusion length and halo function depend on the galactic model, but can be parametrized as
\beq
\lambda_D^2 = 4 \, K_0 \,\tau_E \,\left[\frac{(E/{\rm GeV})^{\delta - 1} - (E^\prime/{\rm GeV})^{\delta - 1}}{1-\delta}\right],
\eeq 
and 
\beq
I(\lambda_D) = a_0 + a_1 {\rm tanh} \left(\frac{b_1 - \ell}{c_1}\right) \left[a_2 \exp \left(- \frac{(\ell - b_2)^2}{c_2}\right) + a_3\right],
\eeq
respectively, with $\ell = \log_{10} \lambda_D / {\rm kpc}$ and the remaining parameters model dependent and listed in Table \ref{tab:galactic-model}.  Our benchmark is the ``MED'' model of propagation and an NFW dark matter profile, though we note that the dark matter profile does not have a significant effect on our results.  We present results using the MIN and MAX models as well.
\begin{table}[hb!]
\begin{center}
\renewcommand{\arraystretch}{1.5}
	\begin{tabular}{c c c c c c c c c c c}
		\hline\hline
		Model & $\delta$ & $K_0~({{\rm kpc}^2/{\rm Myr}})$ & $a_0$ & $a_1$ & $a_2$ & $a_3$ & $b_1$ & $b_2$ & $c_1$ & $c_2$ \\
		\hline
		MIN & 0.55 & 0.00595 & 0.500 & 0.774 & -0.448 & 0.649 & 0.096 & 192.8 & 0.211 & 33.88 \\
		MED & 0.70 & 0.0112 & 0.502 & 0.621 & 0.688 & 0.806 & 0.891 & 0.721 & 0.143 & 0.071 \\
		MAX & 0.46 & 0.0765 & 0.502 & 0.756 & 1.533 & 0.672 & 1.205 & 0.799 & 0.155 & 0.067 \\
		\hline\hline	
	\end{tabular}
	\caption{Parameters used in the fit for galactic propagation for three different propagation models and the NFW dark matter profile as determined in Ref.~\cite{Cirelli:2008id}.}
	\label{tab:galactic-model}
\end{center}
\end{table}

In order to predict the positron fraction, which is the most sensitive observable to the model presented by cosmic ray experiments, we must additionally make a choice about the background electron and positron fluxes.  The positron fraction can be written as
\beq
\frac{\Phi^{e^+}}{\Phi^{e^-} + \Phi^{e^+}} = \frac{\Phi^{e^+}_{\rm sig} + \Phi^{e^+}_{\rm bkg}}{2 \,\Phi^{e^+}_{\rm sig} + \Phi^{e^-}_{\rm bkg} + \Phi^{e^+}_{\rm bkg}} \,, \label{eq:pos-frac}
\eeq
where we have made the assumption, true in our model, that the signal positron and electron fluxes are the same.  Here and where relevant, we use $\Phi$ to represent $d\Phi/dE$ for continuous spectra and the flux in a given bin for binned distributions.  The background positron flux $\Phi^{e^+}_{\rm bkg}$ appearing in the numerator of Eq.~(\ref{eq:pos-frac}) must be taken from a theoretical model, but the background lepton fluxes may be taken either from a model or from data.  We have applied both methods in this work.  For the theoretical model method, we use the background fluxes from Refs.~\cite{Baltz:1998xv,Cirelli:2008id}, given by
\beq
\frac{d\Phi^{e^-}_{\rm bkg}}{dE} = \frac{0.16\, E^{-1.1}}{1+11\, E^{0.9} + 3.2\, E^{2.15}} + \frac{0.70\, E^{0.7}}{1+110\,E^{1.5}+580\, E^{4.2}}  \,,
\eeq
and
\beq
\frac{d\Phi^{e^+}_{\rm bkg}}{dE} = \frac{4.5 \, E^{0.7}}{1+650\, E^{2.3} + 1500\, E^{4.2}} \,,
\eeq
for $E$ in GeV and $d\Phi/dE$ in $ {\rm GeV}^{-1}\,{\rm cm}^{-2}\,{\rm s}^{-1}\,{\rm sr}^{-1}$.
When applying data to determine the sum of electron and positron fluxes, we interpolate the data in Ref.~\cite{Aguilar:2014fea} from AMS-02.

\begin{figure}[th!]
\begin{center}
\includegraphics[width=0.48\textwidth]{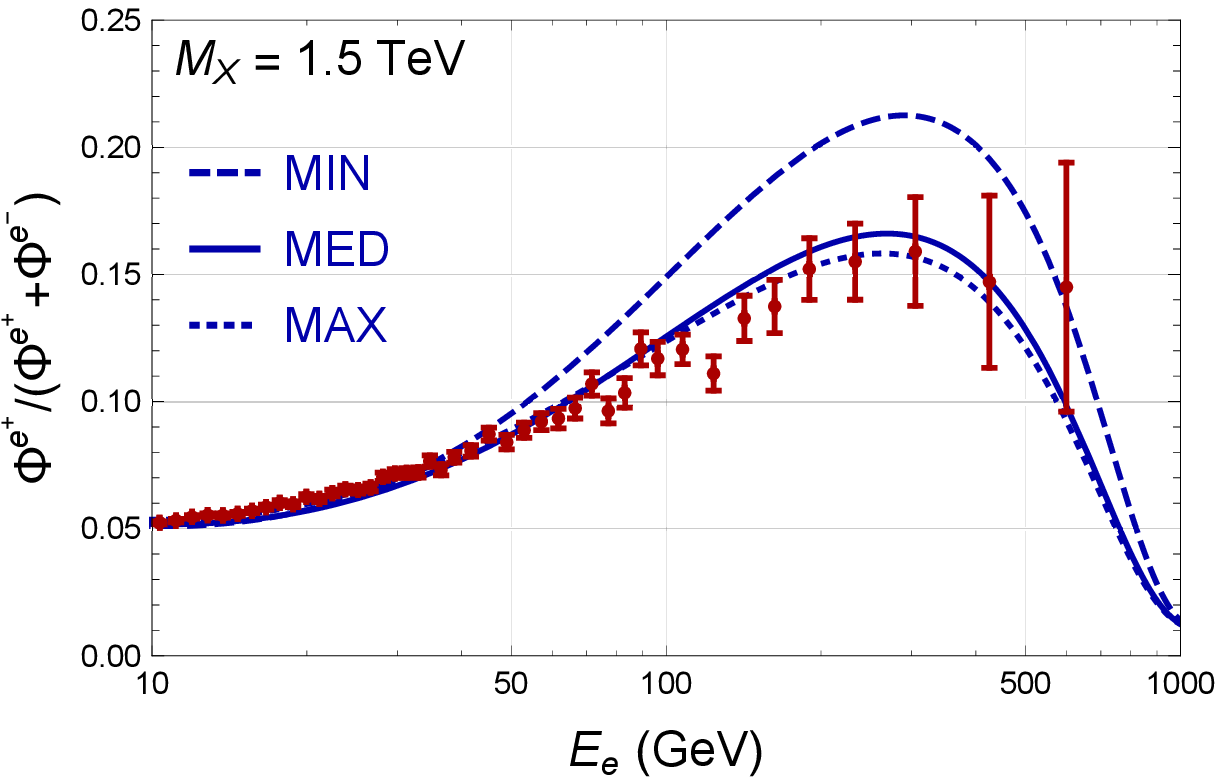}\includegraphics[width=0.48\textwidth]{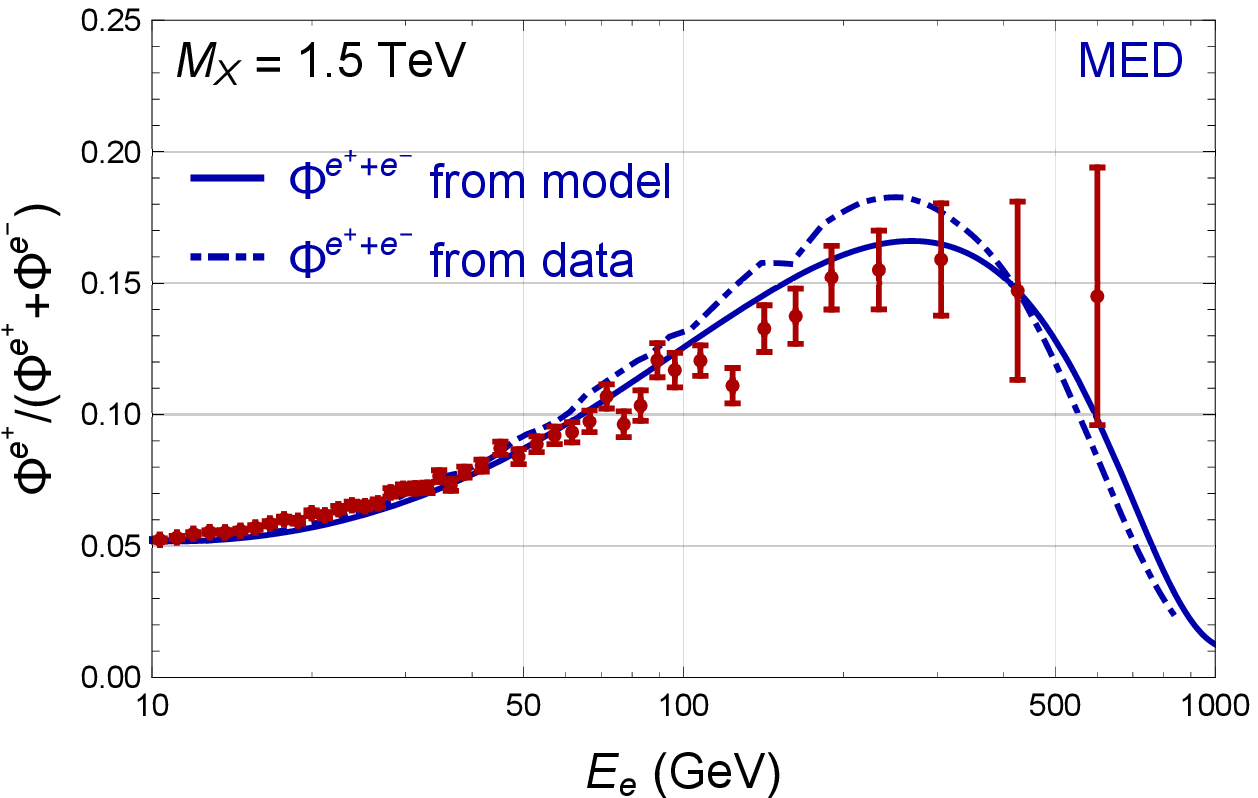}
\caption{Fits to the AMS-02 data with a heavy dark matter state.  All fits assume $M_X = 1.5~{\rm TeV}$ and $M_{\Phi^{\rm D}_s}=0.5$~GeV.  Left: Propagation fit taken from Ref.~\cite{Cirelli:2008id}, using models MIN, MED and MAX with annihilation rates $\langle(\sigma v)_{\rm eff} \rangle = 1.3 \times 10^{-22},~7.9 \times 10^{-23},~7.2 \times 10^{-23}~{\rm cm^3}/{\rm s}$ respectively.  A theoretical flux model is used for the background in all three cases.  Right: MED propagation model with total lepton flux taken from a theoretical model  with $\langle (\sigma v)_{\rm eff} \rangle  = 7.9 \times 10^{-23}~{\rm cm}^3/{\rm s}$ and from the AMS-02 data with $\langle (\sigma v)_{\rm eff}\rangle  = 5.8 \times 10^{-23}~{\rm cm}^3/{\rm s}$.}
\label{fig:ams-fit}
\end{center}
\end{figure}

For a fixed dark matter mass of $M_X = 1.5~{\rm TeV}$, we then determine a cross section that fits the available data.  The data is largely taken from Ref.~\cite{Accardo:2014lma}, though we supplement this data with a new data point presented in Ref.~\cite{Ting:Dec2016Talk}.  The data and the spectrum resulting from our model are shown in Fig.~\ref{fig:ams-fit}. Depending on the propagation model, the required dark matter annihilation rates in our model could vary by a factor of two. For the benchmark ``MED'' propagation model, the required annihilation rate is around $7.9 \times 10^{-23}~{\rm cm^3}/{\rm s}$ and approximately matched by our benchmark model point in Fig.~\ref{fig:sigmav} or \ref{fig:yield}. 

Given this fit to the AMS data, we now turn to the dominant constraint on the models coming from annihilations of dark matter into electromagnetic particles that can distort the spectrum of the CMB radiation.  The relevant annihilations occur just after the time of recombination, at which point the typical dark matter velocity is very small.  The annihilation cross section is thus close to its zero velocity value.  In the SRDM model, there is thus a non-trivial interplay between various constraints.  Annihilations just after recombination must be sufficiently small to not generate CMB distortions, annihilation during freeze-out must be large enough to not overproduce dark matter, and annihilations in the Milky Way today must have the correct rate to generate the observed cosmic ray spectrum.

The PLANCK collaboration has presented general bounds on annihilation of dark matter into electromagnetic particles, {\it i.e.}~electrons and photons~\cite{Ade:2015xua}.  They are sensitive to the rate of energy deposition per unit volume, which can generally be written as
\beq
\label{eq:planck-en-dump}
\frac{dE}{dV dt}(z) = 2\,\tilde{g}\,\rho_{\rm DM}^2(z)\,\frac{f(z)\,\langle \sigma\,v_{\rm rel} \rangle}{m_\chi},
\eeq
where $\tilde{g}$ is a degeneracy factor of $1/2$ for Majorana or real scalar dark matter and $1/4$ for Dirac or complex scalar dark matter; $\rho_{\rm DM}$ is the dark matter energy density at a given redshift $z$; $f(z)$ is an efficiency factor for absorption of electromagnetic energy that we discuss in greater detail below; and $m_\chi$ is the dark matter mass.  We discuss each of these factors in greater detail as they pertain to the model above.

The degeneracy factor $\tilde{g}$ is taken to be $1/2$ for the determination of constraints by the PLANCK collaboration.  This is consistent with the model we work with here, which effectively behaves like Majorana or real scalar dark matter, with only $XX$ annihilation.   Particles annihilate with identical particles, rather than anti-particles.  The dark matter density as a function of $z$ is an observable that is fixed independently.  The factor of $f(z)$ is the efficiency for produced electrons and photons to dump energy into the matter-radiation bath.  At the narrow relevant range of $z = 600$--$1000$, it has been shown \cite{2011PhRvD..84b7302G,Giesen:2012rp,2012PhRvD..85d3522F} that $f$ is nearly independent of $z$, such that a constraint can be determined in terms of a constant $f_{\rm eff}$.  To determine $f_{\rm eff}$ for the model considered here, we follow Ref.~\cite{Slatyer:2015jla}.  In particular, this factor can be determined from Eq.~(2) therein, which we modify to the model above as
\beq
\label{eq:feff-calc}
f_{\rm eff}(M_X) = \frac{\int_0^{3\,M_X/4} dE\,E\,2\,f_{\rm eff}^{e^+e^-}(E)\,dN /dE}{2\,M_X} \,,
\eeq
where $f_{\rm eff}^{e^+e^-}(E)$ is the efficiency for absorption of energy from electrons and positrons and $dN / dE$ is the spectrum of positrons produced in dark matter annihilation given in Eq.~\eqref{eq:spectrum-prod}.  The spectrum $f_{\rm eff}^{e^+e^-}(E)$ is taken from the calculation in Ref.~\cite{Slatyer:2015jla}.   Note once more that the maximum energy that an electron or positron produced in annihilation is $3M_X/4$, unlike in models where two light scalars are produced in annihilation.

The remaining factors in Eq.~\eqref{eq:planck-en-dump} are trivial.  $\langle \sigma v_{\rm rel} \rangle$ should just be taken to be $\langle (\sigma v_{\rm rel})_{\rm eff} \rangle$ in our model, while $m_\chi = M_X$.  The resulting bound from combining the PLANCK collaboration constraint \cite{Ade:2015xua} with Eq.~(\ref{eq:feff-calc}) can be written as
\beq
\langle (\sigma v_{\rm rel})\rangle^{\rm AMS}_{\rm eff}  < 1.5 \times 10^{-23}~{\rm cm}^3 / {\rm s} \times \frac{M_X}{1.5~{\rm TeV}} \,.
\eeq

The constraints above can also be approximately obtained by reading constraints on $VV \to 4\mu$ models with $V$ as a light vector boson or exactly obtained by reading constraints on $\phi\phi \to 4\mu$ models.  All that is required is some rescaling factors that we derive below.  Before beginning, note that the spectrum of electrons in $VV \to 4\mu$ is not very different from those in $\phi\phi \to 4\mu$, leading to similar constraints and predictions for the two models.  The rescaling that we derive here is physically due to two important, but somewhat superficial differences between the SRDM model and the $\phi\phi \to 4\mu$ model.  The first is that the semi-annihilation process in the SRDM model produces only a single scalar that decays to muons, leading to a trivial relative factor of $1/2$.  The second difference has less trivial repercussions: the momentum of the scalar produced in annihilation is $3 M_{\rm DM} /4$ in the SRDM as opposed to $M_{\rm DM}$ in the $\phi\phi \to 4 \mu$ model.  In order to read off constraints on SRDM model using constraints on $\phi\phi \to 4 \mu$ or $VV \to 4\mu$, one needs to rescale the constrained mass $M_{\rm DM}^{\rm eff}$ and $\langle (\sigma v_{\rm rel})\rangle_{\rm eff,4 \mu}$ to the the physical parameters of the SRDM model $M_X$ and $\langle (\sigma v_{\rm rel})\rangle_{\rm eff}$.  We derive this rescaling below.

Since the scalar decaying to muons is produced with fixed momentum in both models, the energy spectrum of positrons (and electrons) is related by a constant factor, so that we have the mapping,
$M_{\rm DM}^{\rm eff} = 3 M_X/4$.  Since the number density of dark matter is given by $\rho_{\rm DM} / M_{\rm DM}$ and $M_{\rm DM} = M_X = 4 M_{\rm DM}^{\rm eff} / 3$, the annihilation rate for SRDM dark matter is suppressed by a factor of $(3/4)^2 = 9/16$, in addition to the factor of $1/2$ described above.  Note that this factor applies to CMB constraints as well, since one explicit factor of $3/4$ is obtained from the factor of $m_\chi$ in the denominator of Eq.~\eqref{eq:planck-en-dump} and second factor of $3/4$ enters in from writing $f_{\rm eff}$ as
\beq
f_{\rm eff} = \frac{3}{4} \frac{\int_0^1 dx  x f_{\rm eff}^{e^+e^-}(x M_{\rm DM}^{\rm eff}) dN/dx}{2},
\eeq
where we change variables to $x = 4 E / 3 M_X = E / M_{\rm DM}^{\rm eff}$.  Combining these two factors, we obtain the relation
\beq
\langle (\sigma v_{\rm rel})\rangle_{{\rm eff},4\mu}  = \frac{9}{32} \, \langle (\sigma v_{\rm rel})\rangle_{\rm eff} \,.
\eeq
This relation allows us to read off any constraint on the annihilation rate of $\phi\phi \to 4 \mu$ directly.

\begin{figure}[th!]
\begin{center}
\includegraphics[width=0.49\textwidth]{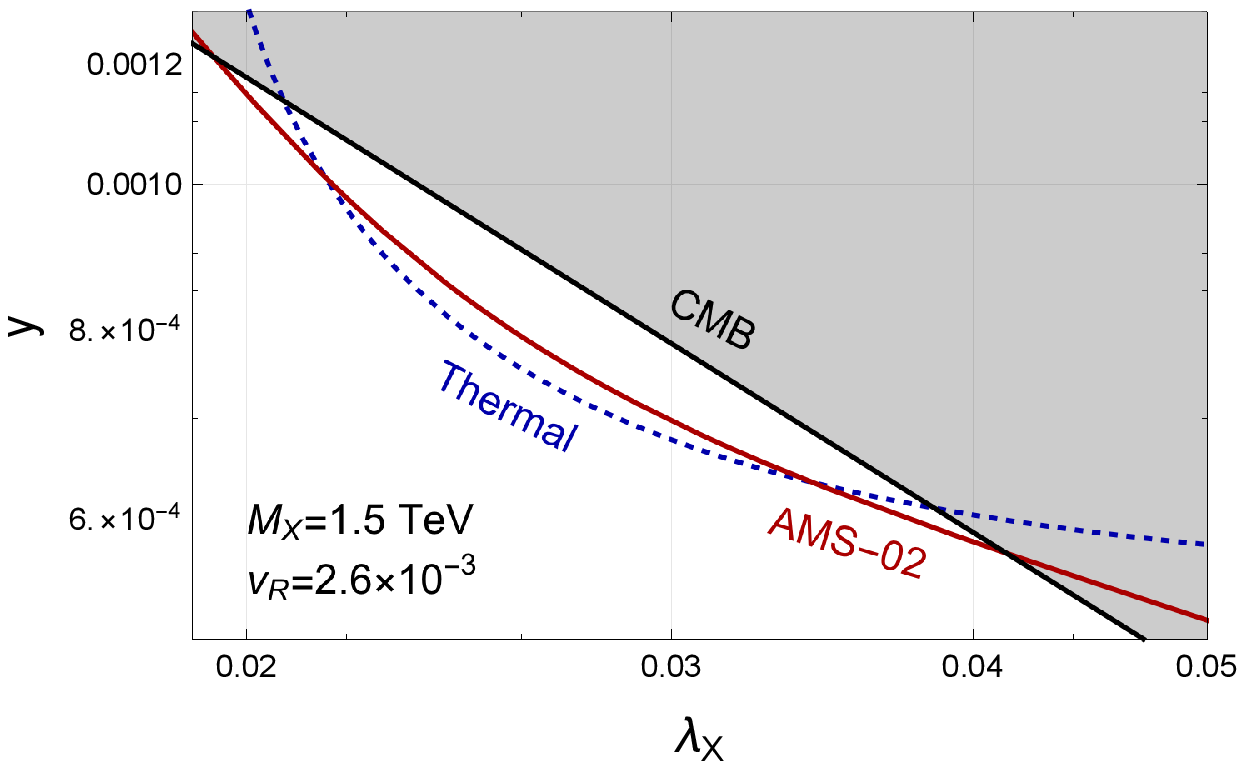} \hspace{3mm}
\includegraphics[width=0.47\textwidth]{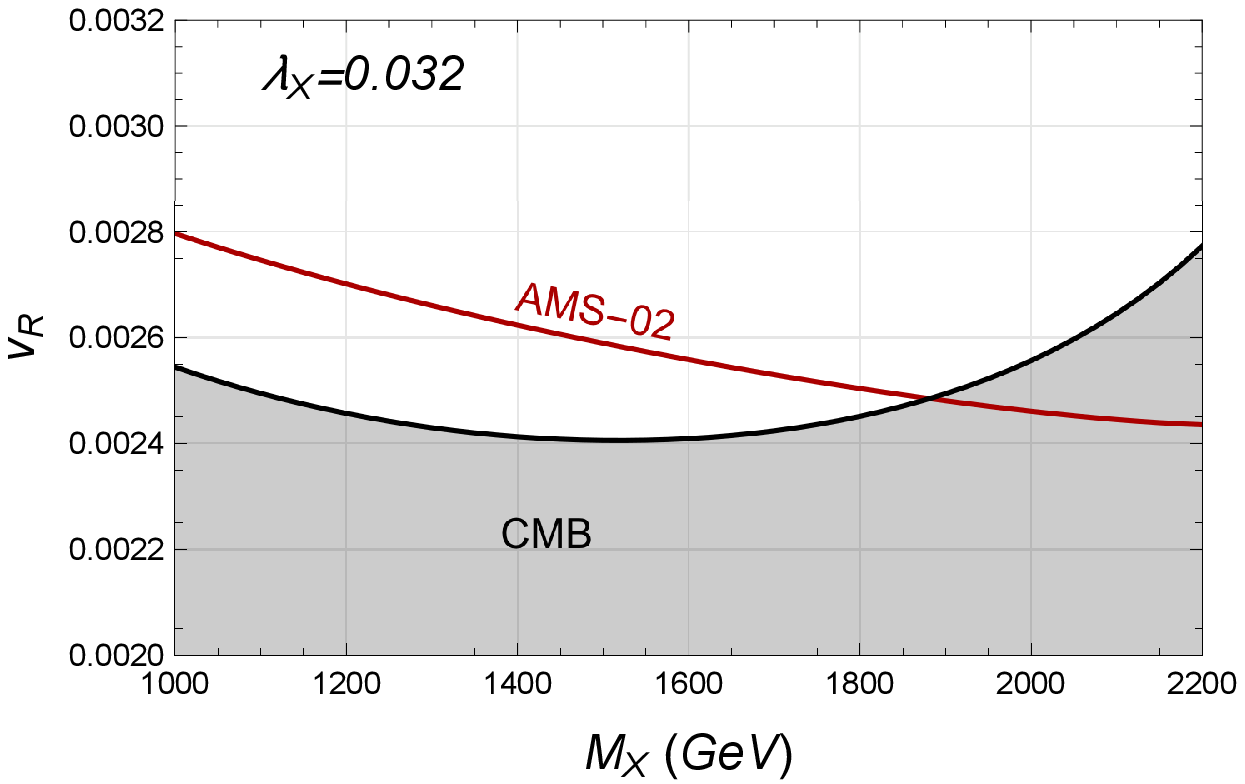}
\caption{Left panel: the parameter space in $\lambda_X$ and $y$ to fit the dark matter thermal abundance and the AMS-02 signal with $\langle \sigma v_{\rm rel}\rangle^{\rm AMS}_{\rm eff} = 7.9\times 10^{-23}\times (M_X/1.5\,\mbox{TeV})^2\,\mbox{cm}^3/\mbox{s}$. The constraint from CMB is approximately taken as $\langle \sigma v_{\rm rel}\rangle^{\rm CMB}_{\rm eff} < 1.5\times 10^{-23}\times (M_X/1.5\,\mbox{TeV})\,\mbox{cm}^3/\mbox{s}$. Right panel: after satisfying the relic abundance via Eq.~(\ref{eq:thermal-relation}), the allowed parameter space to fit the AMS-02 rate and satisfy the CMB constraints.}
\label{fig:model-space}
\end{center}
\end{figure}

Combining the determination of the required parameters to achieve the observed thermal relic abundance, fit the positron fraction data, and evade constraints from annihilations in the recombination epoch, we arrive at results shown in Fig.~\ref{fig:model-space}. It is clear from this figure that although our model can satisfy the constraints from CMB, the allowed parameter space is not that large. Given the uncertainties of the cosmic ray propagation model and preferred annihilation rates for positron excess, we will not perform global fit to all experimental data to search for the allowed model parameter space. In the right panel of Fig.~\ref{fig:model-space} and fixing the coupling $\lambda_X = 0.032$, we can see that to explain the AMS-02 preferred annihilation rate, the dark matter mass has to be below around 1.9 TeV. This is another interesting general feature of the resonant dark matter model, in which an upper bound on dark matter mass exists.

\section{Additional Signals}
\label{sec:add-signals}
The main signal of the SRDM model is annihilation that produces muons and thereby cosmic ray positrons.  In this section, we discuss other possible mechanisms for discovering dark matter.  As usual, there are three possible approaches for confirming the SRDM model: direct detection, indirect detection, and collider production.  We discuss these in turn.

Direct detection of the SRDM model is challenging.  The light Goldstone modes do not couple directly to the dark matter state without sufficient energy to excite the heavy $X^{\rm R}$ states and can only mediate interactions with the proton at one loop, while the heavy $\Phi$ modes lead to a large mass suppression of their interaction cross section.   
The dominated heavy $\Phi$ mediated tree-level interaction, on the other hand, has an estimated cross section of
\beq
\sigma_{\Phi_{\rm T},\Phi_8} \sim \frac{1}{16\, \pi \, M_X^2} \, \epsilon_2^2 \, y^2  \, \frac{M_p^4}{M_h^4} \sim 10^{-58}~{\rm cm}^2 \times \left(\frac{1.5~{\rm TeV}}{M_X}\right)^2 \, \left(\frac{\epsilon_2}{10^{-4}}\right)^2\,\left(\frac{y}{10^{-3}}\right)^2 \,, 
\eeq
which is too small to be accessed in the near future. 

Indirect detection of positrons and gamma rays from the SRDM model has been discussed above, but as seen in Fig.~\ref{fig:feynman}, another SM product of dark matter annihilation is neutrinos.  The neutrinos are emitted with an energy $E \sim \delta m \sim 1~{\rm MeV}$.  Such neutrinos are relatively low energy and are difficult to detect without an enormous flux, such as that from the Sun.  To detect such neutrinos, detectors use either inverse beta decay or scattering off electrons.  In either case, the cross section is determined by the four Fermi weak interaction and is thus highly suppressed.  In addition, the flux is expected to be of order
\beq
\Phi_\nu \sim J \, \rho_{\odot}^2 \, r_{\odot} \, \frac{\langle(\sigma v)_{\rm eff}\rangle}{M_X^2}  \sim 10^{-11}~{\rm cm}^{-2} \, {\rm s}^{-1} \times J \,,
\eeq
where $J$ is a dimensionless line-of-sight integral that is of order $100$--$1000$ for the dominant contribution from the galactic center.  For comparison, the flux of solar neutrinos near $E = 1~{\rm MeV}$ is ${\rm few} \times 10^8~{\rm cm}^{-2} \, {\rm s}^{-1}$ \cite{Bahcall:2004pz}. The predicted MeV neutrino flux in our model is far too small to be detected with current detectors. 

\begin{figure}[th!]
\begin{center}
\includegraphics[width=0.48\textwidth]{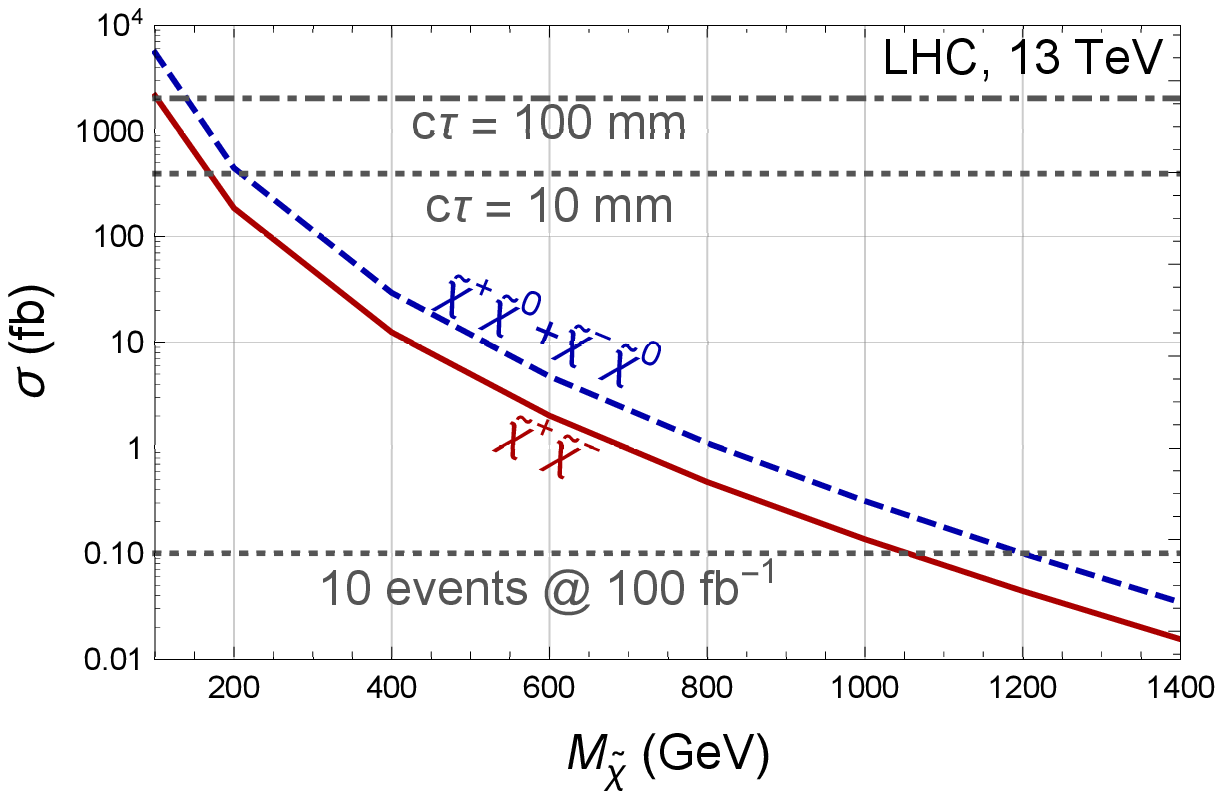} \hspace{3mm}
\includegraphics[width=0.48\textwidth]{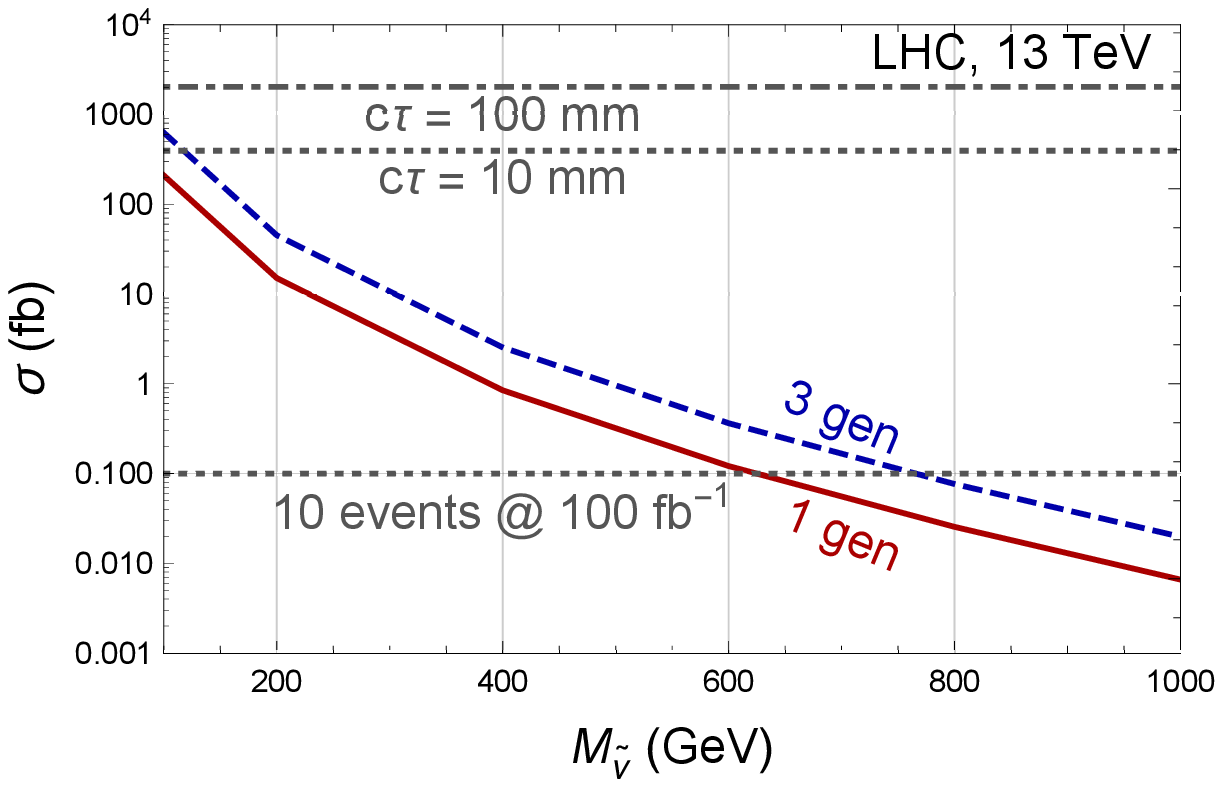}
\caption{Left panel: Production cross section for the dominant Higgsino states when the Higgsino is the lightest SM partner.  We show the channels $\tilde{\chi}^+ \tilde{\chi}^-$ (solid red) and $\tilde{\chi}^\pm \tilde{\chi}^0$ (dashed blue).  Right panel: Production cross section for sneutrinos when the sneutrino is the lightest SM partner.  We present cases where there is only one light sneutrino state (solid red) and where there are three degenerate light sneutrino states (dashed blue).  Current estimated constraints from an ATLAS search for displaced lepton tracks~\cite{ATLAS:2016jza} are shown for lifetimes of $100~{\rm mm}$ (dot dashed gray) and $10~{\rm mm}$ (dotted gray).}
\label{fig:lhc-xsec}
\end{center}
\end{figure}

The most promising additional means of detecting the SRDM model is by producing states at colliders.  Our model is by necessity added onto a supersymmetrized SM.  Since the Goldstone multiplet states are rather light, with masses $\lesssim 1~{\rm GeV}$, the lightest SM superpartner states can decay to them if other R-parity violating decay modes are subdominant.  If this decay is sufficiently fast, then the SRDM leads to a striking prediction of several muon jets arising from the collimated muons in the Goldstone scalar decays that end the decay chain~\cite{ArkaniHamed:2008qp,Baumgart:2009tn,Bai:2009it,Falkowski:2010gv}. 
\begin{figure}[th!]
\begin{center}
\includegraphics[width=0.55\textwidth]{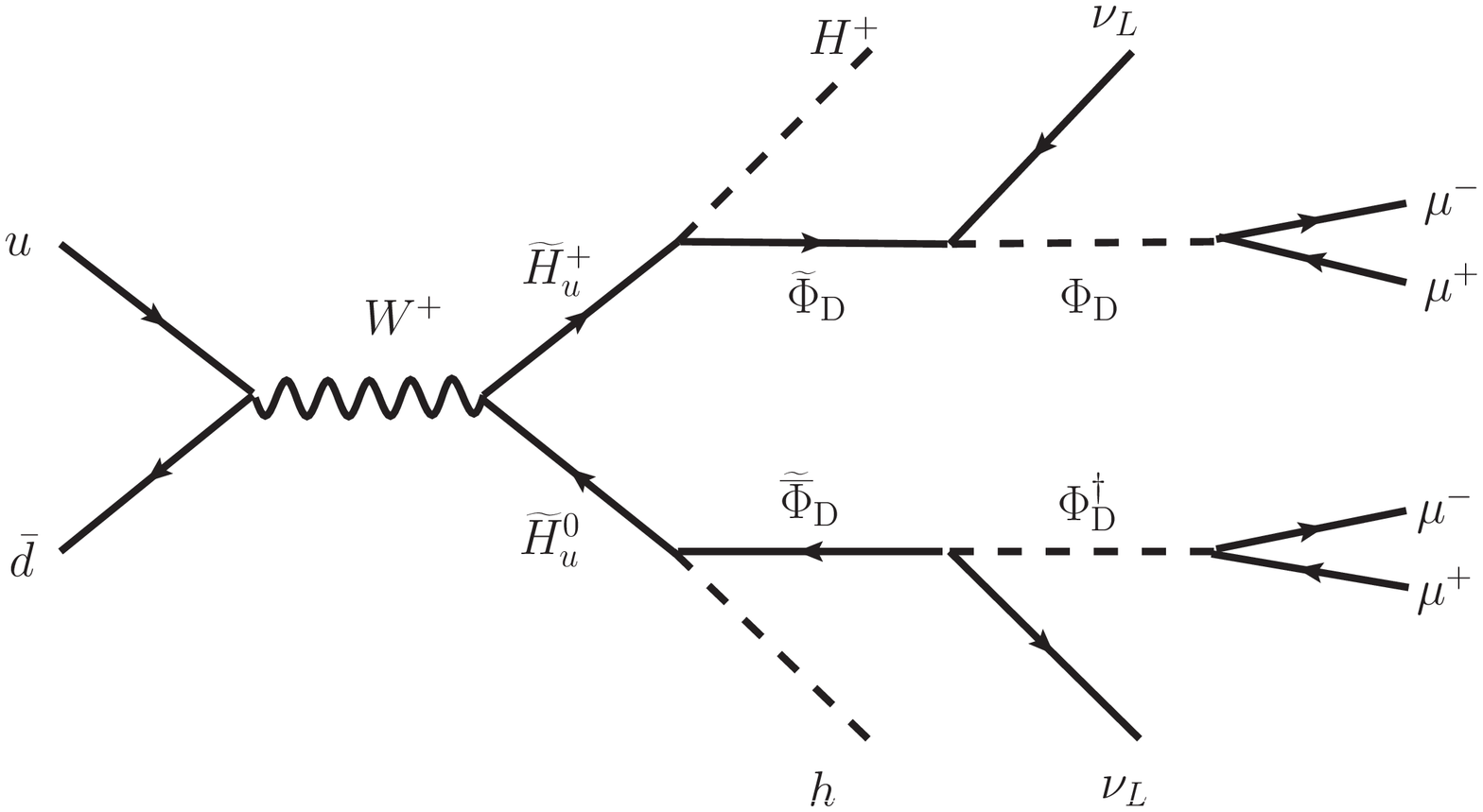}
\caption{A representative Feynman diagram for the collider signatures with Higgs fields and collimated (displaced) ``dimuon jet'' in the final state.}
\label{fig:collider-feynman}
\end{center}
\end{figure}
The mediation to the dark sector can be achieved from production of Higgsino, charginos or sneutrinos. We therefore consider the minimal scenario in which either a Higgsino or a sneutrino is the lightest SM partner.  Any spectrum in which there are light squarks or gluinos will of course be easier to detect. In Fig.~\ref{fig:collider-feynman}, we show a representative Feynman diagram for the signature at the LHC.  Higgsino states are dominantly pair produced via $s$-channel $\gamma^*$, $Z$, and $W$.  The chargino states can be produced as well, assuming they are nearly degenerate with the neutralino states. The dominant production modes are $\tilde{\chi}^\pm \tilde{\chi}^0$ and $\tilde{\chi}^+ \tilde{\chi}^-$, as the $\tilde{\chi}^0 \tilde{\chi}^0$ only couples via the $Z$ and accidental cancellation suppresses this channel by a factor of a few.  Depending on the cutoff of the higher-dimensional operator $\Phi_{\rm D} \overline{\Phi}_{\rm D} H_u L/\Lambda_\Phi$ operator in Eq.~\eqref{eq:higher-dim} and the coefficient of the SUSY Higgs-potal operator $\epsilon_2 \Phi_{\rm D} H_u H_d$, the neutralino could decay into the SM Higgs plus $\Phi_{\rm D} \nu$ and the chargino could decay into charged Higgs plus $\Phi_{\rm D} \nu$, or they can also decay to multiple PNGB's as  $\Phi_{\rm D} \Phi_{\rm D} \nu$ and $\Phi_{\rm D} \Phi_{\rm D} \ell$. For the first possibility with the dominant decay for Higgsino as $h \tilde{\Phi}_{\rm D}$ (ignoring the soft neutrino), one has two collimated muons along with a Higgs that together reconstruct the Higgsino mass. The production cross section for the dominant modes is shown in the left panel of Fig.~\ref{fig:lhc-xsec}. Because of the striking signature properties, we anticipate a very good coverage of the model parameter space at the LHC Run 2. 

The other minimal scenario, with only light sneutrinos, has a smaller cross section both due to the fact that the sneutrinos don't carry charge and that they are scalars.  The left-handed sleptons corresponding to the relatively light sneutrinos may also be light, but we neglect this possibility here.  The dominant decay also occurs via the operator $\Phi_{\rm D} \overline{\Phi}_{\rm D} H_u L$ in Eq.~\eqref{eq:higher-dim}.  The decay products are $\Phi_{\rm D} \tilde{\Phi}_{\rm D}$. The production cross section is shown in the right panel of Fig.~\ref{fig:lhc-xsec}.

In either of the above cases, the final state may contain four $\Phi_{\rm D}$ PNGB's, each of which decays to a collimated pair of muons since the $\Phi_{\rm D}$'s are produced at highly relativistic energies.  Therefore, the signal of these models is multiple muon jets, along with other possible SM particles.  The upper bound on $\Lambda_\Phi$ in the SRDM model is very mild and arises by requiring that the heavier states in the $\Phi_{\rm D}$ supermultiplet decay sufficiently promptly on a cosmological.  Their decays to $\Phi_{\rm D}$ therefore may or may not be prompt in colliders, depending on the value of $\Lambda_\Phi$.  The $\Phi_{\rm D}$, on the other hand, has a lifetime of order $10^{-11}~{\rm s}$, as seen in Eq.~\eqref{eq:goldstone-decay}.  The decays of $\Phi_{\rm D}$ are therefore generally displaced by $\mathcal{O}(3~{\rm mm})$ at rest and $\mathcal{O}(30~{\rm cm})$ with a Lorentz boost of  $\mathcal{O}(100)$.

CMS and ATLAS have searched for long-lived scalars having displaced decays to muons at 7 TeV~\cite{CMS:2014hka} and 13 TeV~\cite{ATLAS:2016jza}, respectively.  The ATLAS search has greater sensitivity to the SRDM model at the moment.  Since there is some flexibility in the lifetime of the $\Phi_{\rm D}$ states, we indicate two benchmark lifetimes of $c\tau = 10~{\rm mm}$ and $c\tau = 100~{\rm mm}$ in Fig.~\ref{fig:lhc-xsec}.  These bounds should be interpreted as rough, since the models considered in Ref.~\cite{ATLAS:2016jza} do not map exactly onto the model we consider.  We focus on the FRVZ~\cite{Falkowski:2010gv} $4\gamma_d$ model with $m_H = 800~{\rm GeV}$ and $m_{\gamma_d} = 400~{\rm MeV}$ since this is the model that most closely matches the interesting portion of parameter space of SRDM model.  We note that the LHC is just beginning to have sufficient data to probe these processes in the most difficult channels. For our SRDM model, because of the additional energetic SM particles in the final state, both the trigger choices and reduction of backgrounds in our model are much easier than the FRVZ model. A more careful collider study is needed to know the final reach of neutralino or sneutrino masses at the LHC Run 2, which we leave to future study. 

\section{Discussion and Conclusions}
\label{sec:conclusion}

There are some effects we have neglected up to this point in our calculations that should have a minimal effect.  We have thus far neglected the $SU(3)_f$-breaking effects in the superpotential that will feed back at loop-level into the K\"ahler potential, which could cause small distortions of the delicate factor of in the $X$ mass spectrum required for the resonant enhancement of the semi-annihilation cross section.  The largest of these effects comes from the $\lambda_X$ interaction in Eq.~(\ref{eq:superpotential-breaking}).  The one-loop contribution to  the wavefunction renormalization of $X$ in the K\"ahler potential  is of order $\lambda_X^2 / (16 \pi^2) \sim 10^{-6}$, leading to a contribution to $\delta m / M_X \sim 10^{-6}$, comparable to the benchmark discussed above.  We note this additional contribution, but it is not an issue for the phenomenological analysis above.  Other contributions to the mass splitting are suppressed smaller parameters, higher scales, or more loops and are negligible.

Another loop-level effect on the spectrum affects the SM particle properties.  The first term in the decay superpotential of Eq.~\eqref{eq:higher-dim} can generate a Majorana neutrino mass.  The contribution is of order
\beq
\Delta m_\nu \sim \frac{v_u^2\, M_X\,a_{ii}^2 }{16 \pi^2 \Lambda_X^2} \,\frac{\delta m}{M_X}\sim 0.06~{\rm eV} \times \left(\frac{10^5~{\rm GeV}}{\Lambda_X/a_{ii}}\right)^2 \, \left(\frac{\delta m/M_X}{10^{-6}}\right)\, \left(\frac{M_X}{1.5~{\rm TeV}}\right) \,.
\eeq
Note that the generation of Majorana neutrino masses necessarily requires SUSY breaking by holomorphy, so additional suppression factor of $\delta m/M_X$ is anticipated.  A lower bound is set on $\Lambda_X$, since the largest possible neutrino mass is around $0.23~{\rm eV}$ from cosmological constraints \cite{Ade:2015xua} and is around $0.31~{\rm eV}$ for Majorana effective mass determined by neutrinoless double beta decay ($0\nu\beta\beta$) \cite{Guzowski:2015saa}.  Since the mass generated is in fact a Majorana mass, the constraint from $0\nu\beta\beta$ should be taken into account and we can set a constraint of $\Lambda_X/a_{ii} \gtrsim 5 \times 10^4~{\rm GeV}$. We also note that the dark flavor-changing operator with an coefficient of $a_{12}$ will not generate neutrino Majorana mass by itself. This is because the lepton-number is conserved if other coefficients $a_{11}$, and $a_{22}$ are tiny.  

One final effect to consider is the feedback of SM SUSY breaking into the dark sector.  The dominant mediation of SUSY breaking occurs via the SUSY Higgs portal parameter $\epsilon_2 \lesssim 10^{-4}$.  All such effects on the spectrum are highly suppressed as they require a double insertion of operators coupling to the SM.  Loop effects from SUSY breaking could also in principle destabilize the desired vacuum alignment which is dominantly along the $\Phi^8$ direction. Such effects should only led to small corrections that could not induce a large change to a completely different vacuum structure.

In addition to these higher order effects on the spectrum, we have made a few assumptions in our solution to the Boltzmann equations.  The semi-annihilation process generates a quasi-relativistic $X$ or $\overline{X}$ in the final state.  We have made the assumption that there is an interaction that allows these relativistic dark matter particles to rapidly thermalize with the non-relativistic dark matter population.  This interaction also erases any distortions caused by the non-trivial dark matter momentum dependence of the semi-annihilation cross section on the non-relativistic dark matter population. Independent of the interpretation of the AMS-02 positron excess, the velocity-dependent semi-annihilation scenario introduced in our model may have other interesting implications on the large scale structure with a mixture of cold and warm dark matter components.  

Before concluding, we note that the precise predictions for the spectrum and flux of cosmic ray positrons is quite sensitive to the details of the electron and positron propagation in the galaxy, the size of the standard background of positrons and the velocity distribution of dark matter in the galactic halo. We have shown a few examples of how the spectrum could change due to the first of these. One particular effect is on the upper bound on the allowed dark matter mass to accommodate all the data fit in this work. This uncertainty could leave a larger window of viability for the SRDM model.

In summary, we have developed and studied a SRDM model where annihilations producing positrons are resonantly enhanced in the Milky Way. The model evades strong constraints from precision CMB measurements, while explaining the observed dark matter relic abundance and high energy cosmic ray positron excess. The resonance must have a mass very close to twice the dark matter mass, which is achieved by spontaneously breaking an $SU(3)_f$ flavor symmetry to $SU(2)_f \times U(1)_f$ under which dark matter is charged. The factor of two relation in the VEV-induced mass for a triplet of $SU(3)_f$ provides a novel mechanism for naturally explaining resonant dark matter. The PNGB of spontaneous flavor symmetry breaking decay dominantly to muons, leading to an implementation of a class of models where dark matter annihilates into particles that only later decay to the SM. Such models are known to be far safer from gamma ray constraints at the large cross sections required to explain the positron excess. Since the model structure requires supersymmetry for stability, weak scale superpartners of the SM lead to distinctive signals for colliders, such as decay chains with an SM Higgs and collimated muons. We have shown that searches for displaced lepton jets could have sensitivity to such models, while searches more optimized for the particular structure of this model could increase sensitivity at the LHC.

\subsection*{Acknowledgments}
We would like to thank Spencer Chang, Daniel Chung, Marco Cirelli, Lisa Everett, Katherine Freese, Ian Low and Lian-Tao Wang for useful discussion. This work is supported by the U. S. Department of Energy under the contract DE-FG-02-95ER40896.

\appendix

\section{General Parametrization}
\label{sec:general-parametrization}
Starting from the generalized parametrization, we have the effective annihilation rate as
\beqa
(\sigma v_{\rm rel})_{\rm eff} = \sigma_0 \, \frac{\delta^2 + \gamma^2}{(\delta - v_{\rm rel}^2/4)^2 + \gamma^2} \,.
\eeqa
Here, the parameter $\gamma$ is treated to be independent of $v_{\rm rel}$. If only the complex scalar $X_p^1$ is the dark matter, the degrees of freedom is $g=2$. To suppress the prediction for the CMB while explain the rate for the AMS-02, we need to work in the the limit of $\delta \gg \gamma$. 

To satisfy the relic abundance, $\Omega h^2 = 0.112$, the following simple relation is needed 
\beqa
\frac{\Omega h^2}{0.112} = 7.4\times 10^{-11} \, \frac{\gamma}{\sigma_0 \, \delta^2}  \Rightarrow
\frac{\gamma}{\sigma_0 \,\delta^2} = 1.35 \times 10^{10}\, \mbox{GeV}^2 \,. 
\label{eq:thermal}
\eeqa

For a narrow resonance and if the pole can be reached for the integration, the annihilation rate is
\beqa
(\sigma v_{\rm rel})_{\rm eff} = \frac{\pi\,\sigma_0 \, \delta^{3/2}}{\gamma} \, \delta( v_{\rm rel} - 2 \sqrt{\delta}) \,.
\eeqa
So, the prediction for AMS-02 is 
\beqa
\langle \sigma v_{\rm rel}\rangle^{\rm AMS}_{\rm eff} = \frac{4\pi\,\sigma_0 \, \delta^{5/2}}{\gamma} \, \frac{4\pi}{(2\pi v_0^2)^{3/2}} \, e^{- 2 \delta/v_0^2} \,.
\label{eq:AMS-general}
\eeqa
\begin{figure}[th!]
\begin{center}
\includegraphics[width=0.48\textwidth]{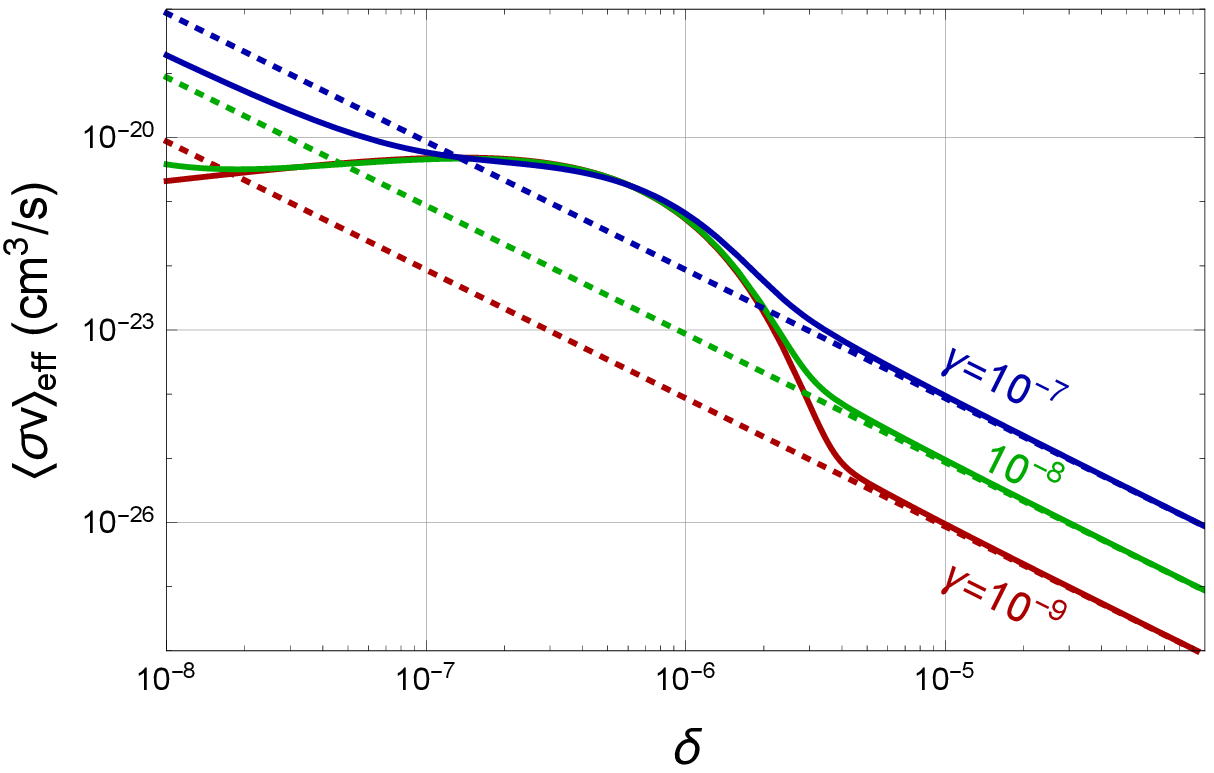} \hspace{3mm}
\includegraphics[width=0.48\textwidth]{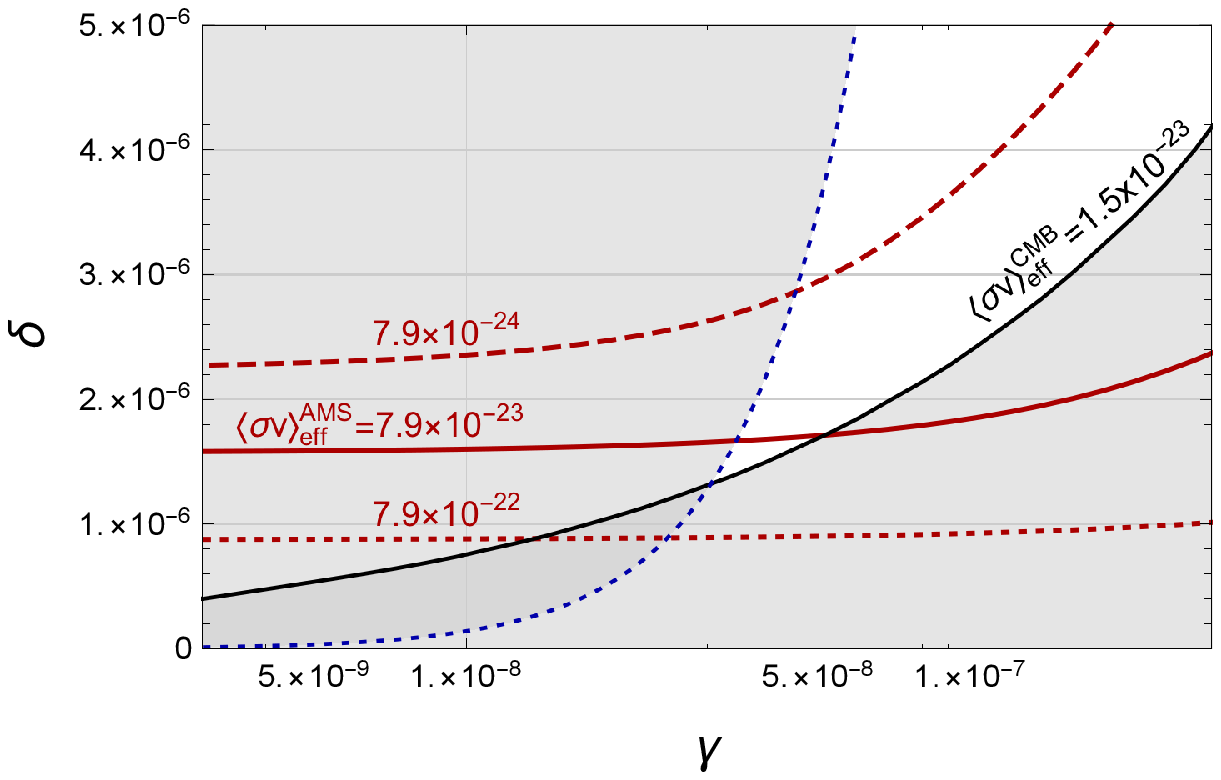}
\caption{Predicted cross sections (solid lines) for annihilation in the Milky Way (as would be seen by AMS-02) and during the recombination era (dotted lines), which lead to CMB constraints.  The low velocity cross section is fixed by requiring that the model achieves the observed present day dark matter abundance, while $M_X$ is fixed to $1.5~{\rm TeV}$ in order to obtain a good fit to the AMS-02 data.  The AMS-02 data fit above had a best fit cross section of $\langle \sigma v\rangle_{\rm eff}^{\rm AMS}  = 7.9 \times 10^{-23}~{\rm cm}^3 / {\rm s}$, while the constraint from annihilations during recombination on $M_X = 1.5~{\rm TeV}$ dark matter is $\langle\sigma v \rangle_{\rm eff}^{\rm CMB} < 1.5 \times 10^{-23}~{\rm cm}^3 / {\rm s}$. The blue  dotted line is the additional constraint on the SRDM model as in Eq.~(\ref{eq:bound-delta-gamma}).}
\label{fig:sigmav-AMS}
\end{center}
\end{figure}

For our case, we have the effective annihilation rate as
\beqa
(\sigma v_{\rm rel})_{\rm eff} =  \frac{3\,y^2 \lambda_X^2}{4\,\pi\,M_X^2\,\left[ (v_{\rm rel}^2 - v_{\rm R}^2)^2 + \frac{1}{64\pi^2} (3\,y^2 + \lambda_X^2 v_{\rm rel})^2 \right]} \,.
\label{eq:breit-wigner-2}
\eeqa
In the limit of $\gamma \ll \delta$, we have the following match relations
\beqa
\sigma_0 \, \delta^2 = \frac{3\,y^2 \lambda_X^2}{64\pi\,M_X^2}  \,, \qquad 
\gamma = \frac{1}{32\pi}\left(3\,y^2 + \lambda_X^2 v_{\rm rel}\right) \,, \qquad \delta = v_{\rm R}^2/4 \,.
\eeqa
To simplify our discussion, we first treat the $v_{\rm rel}$ in $\gamma$ as a constant with $v_{\rm rel} \approx v_{\rm R} = 2 \sqrt{\delta}$. In general, a larger value of $\gamma$ (allowed by CMB constraints) can make our model easier to fit. This is because of the simple math relation $2 a b \leq a^2 + b^2$. Defining the parameter $\kappa = 1.35 \times 10^{10}$~GeV$^2$ from Eq.~(\ref{eq:thermal}), the condition to have real solutions for $y$ and $\lambda_X$ is 
\beqa
\delta  < \frac{4\,\pi^2\,\kappa^2}{M_X^4}\, \gamma^2 \,.
\label{eq:bound-delta-gamma}
\eeqa
For $M_X=1.5$~TeV and $\gamma = 3.5\times 10^{-8}$, one needs $\delta < 1.74 \times 10^{-6}$. Let's choose $\gamma = 3.5\times 10^{-8}$ and $\delta = 1.67 \times 10^{-6}$, the corresponding model parameters are $v_{\rm R} = 2.6\times 10^{-3}$ and $(y, \lambda_X) = (7.1\times 10^{-4}, 2.8\times 10^{-2})$ or $(8.2\times 10^{-4}, 2.4\times 10^{-2})$. For this model point, we have $\langle \sigma v_{\rm rel}\rangle^{\rm AMS}_{\rm eff} = 7.9\times 10^{-23}\,\mbox{cm}^3/\mbox{s}$ and $\langle \sigma v_{\rm rel}\rangle^{\rm CMB}_{\rm eff} = 1.1\times 10^{-23}\,\mbox{cm}^3/\mbox{s}$.

Finally, we also note that there is an upper bound on the dark matter mass in our model. From Eq.~(\ref{eq:AMS-general}), we can rewrite the annihilation rate for AMS-02 as 
\beqa
\langle \sigma v_{\rm rel}\rangle^{\rm AMS}_{\rm eff}  = \frac{\eta(\delta, v_0)}{\kappa} \,, \qquad \qquad \mbox{with}\quad
\eta(\delta, v_0) \equiv \delta^{1/2} \, \frac{16\pi^2}{(2\pi v_0^2)^{3/2}} e^{- 2 \delta/v_0^2} \,.
\eeqa
In this general parametrization, the prediction for the annihilation rate in the CMB era is simply $\langle \sigma v_{\rm rel}\rangle^{\rm CMB}_{\rm eff}  = \sigma_0 = \gamma/(\kappa\,\delta^2)$. Requiring an upper bound on the ratio of those two annihilation rates, we have
\beqa
\frac{\langle \sigma v_{\rm rel}\rangle^{\rm CMB}_{\rm eff} }{\langle \sigma v_{\rm rel}\rangle^{\rm AMS}_{\rm eff} } \leq R^{\rm max}\times\, \left( \frac{1.5~\mbox{TeV}}{M_X} \right) \Rightarrow  \gamma \leq R^{\rm max}\, \delta^2 \, \eta\,\times\, \left( \frac{1.5~\mbox{TeV}}{M_X} \right)  \,,
\label{eq:bound-gamma-eta}
\eeqa
where numerically we have $R^{\rm max}\approx 0.19$ from Section~\ref{sec:ams-02}. Combining the two inequalities in Eqs.~(\ref{eq:bound-delta-gamma}, \ref{eq:bound-gamma-eta}), we arrive at the following upper bound on the dark matter mass
\beqa
M_X \leq \left[2\,\pi \kappa\,R^{\rm max}\, \delta^{3/2}\,\eta(\delta, v_0)  \right]^{1/3}\times(1.5~\mbox{TeV})^{1/3} \approx 1.5~\mbox{TeV} \,,
\eeqa
after we use $\delta = 1.67\times 10^{-6}$ to fit the rate for AMS-02. There is another weaker bound from satisfying the narrow-width condition of $\gamma < \delta$ and Eq.~(\ref{eq:bound-delta-gamma}).

\section{Details of Boltzmann Equation and Kinetic Decoupling}
\label{sec:kinetic-functions}
In order to study kinetic decoupling, additional moments of the full Boltzmann equation are required. It is conventional and convenient to parametrize the kinetic coupling by the variable $y$ defined as
\beq
y = \frac{1}{s^{2/3} n} \int \frac{d^3p}{(2\pi)^3} \mathbf{p}^2 f(p),
\eeq
where $f$ is the phase space distribution of $X_p^1$. The variable $y$ is chosen such it goes to a constant after kinetic decoupling and such that in kinetic equilibrium with the SM bath $y_{\rm EQ} = 3 \, M_X^2 \, x / s^{2/3}(x = 1)$.	The contribution to the evolution of $y$ due to elastic scattering off of other species, namely neutrinos in the SRDM model, has been studied in Ref.~\cite{Bringmann:2006mu}, where it is found that
\beq
\label{eq:boltzmann-elastic}
\left.\frac{dy}{dx}\right|_{\rm elas} = - \frac{1}{H\,x} \, 2\, M_X\, c(T) \, (y - y_{\rm EQ}),
\eeq
where $c(T)$ is given by
\beqa
c(T) &=& \frac{1}{12(2\pi)^3\,M_X^4\, T} \int dk\,k^4\,(e^{k/T} + 1)^{-1}\,\left[1- (e^{k/T} + 1)^{-1}\right] \, \sum_f \, \widetilde{|{\cal M}|^2},
\eeqa
for scattering off of massless fermions $f$. Here, $k$ is the momentum of the relativistic fermion. We define the Mandelstam $t$ averaged amplitude by
\beq
\label{eq:t-average}
\widetilde{|{\cal M}|^2} = \frac{1}{8\,k^4} \int_{-4\,k^2}^0 |{\cal M}|^2 (-t) dt \,.
\eeq
For scattering off neutrinos in the SRDM model, there are two different relevant amplitudes.  If the incoming state has a neutrino (anti-neutrino) and the final state has a neutrino (anti-neutrino), then the amplitude is given by
\beq
|\mathcal{M}|^2 = \frac{\lambda_\nu^4\, [(s - M_X^2)^2 + s\,t]}{(s - M_{\tilde{X}^2}^2)^2 + \Gamma_{\tilde{X}^2}^2(s) \, s}.
\eeq
If the incoming state has a neutrino (anti-neutrino) and the final state has an anti-neutrino (neutrino), then the amplitude is given by
\beq
|\mathcal{M}|^2 = \frac{\lambda_\nu^4\, M_{\tilde{X}^2}^2 \, (-t)}{(s - M_{\tilde{X}^2}^2)^2 + \Gamma_{\tilde{X}^2}^2(s) \, s}.
\eeq
Here, we define $\lambda_\nu = a_{12} v_u / \sqrt{2} \Lambda_X$.  After integrating over $t$ as prescribed by Eq.~\eqref{eq:t-average}, summing over all four possible combinations of neutrinos, and using $s = M_X^2 + 2 \, M_X \, k$, we find, to leading order in $k / M_X$ and  $\delta_2 = \delta m / M_X$,
\beq
\sum_f \widetilde{|{\cal M}|^2} = \frac{2 \, \lambda_\nu^4 \, k^2}{(k - \delta m)^2 + \Gamma_{\tilde{X}^2}^2 / 4}.
\eeq
This contribution tends to push the dark sector toward equilibrium with the relativistic fermion bath off of which it is scattering. There is an additional contribution in the model considered here due to the semi-annihilation process. To leading order in $1/x$, the dominant contribution to the evolution of $y$ is due to the large momentum of the final state $X_1^p$ state, which is produced with a momentum
\beq
\mathbf{p}_{X,{\rm out}}^2 = \frac{9 M_X^2}{16} + \mathcal{O}(\mathbf{p}_{X,in}^2),
\eeq
where the additional contributions go like the very non-relativistic momentum of the incoming $X$ particles, which is suppressed by $1 / x$. By integrating the Boltzmann equation weighted by $\mathbf{p}^2$, the contribution to $y$ of this injection is then given by
\beq
\label{eq:boltzmann-semi-ann}
\left.\frac{dy}{dx}\right|_{\rm semi-ann} = \frac{9 \, M_X^2 \, s^{1/3}}{32\, H\, x} \, \langle \sigma v_{\rm rel} \rangle_{\rm eff, x_{\rm DM}} \, Y  \,.
\eeq

\begin{figure}[th!]
\begin{center}
\includegraphics[width=0.48\textwidth]{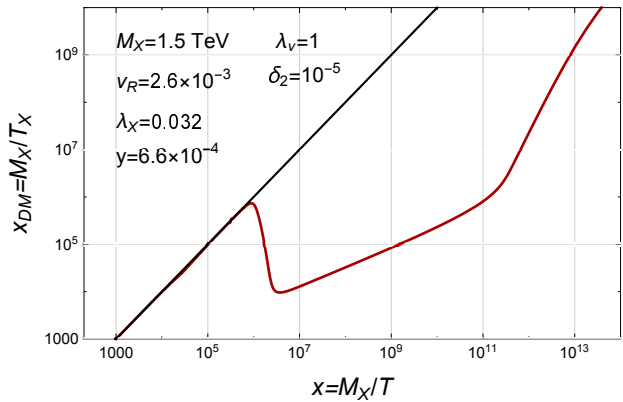} 
\caption{The dark matter temperature parameter as a function of the SM temperature parameter.}
\label{fig:xdm}
\end{center}
\end{figure}

Our full solution for the abundance of dark matter is obtained by solving the coupled Eqs.~\eqref{eq:Y-equation}, \eqref{eq:boltzmann-elastic}, and \eqref{eq:boltzmann-semi-ann} under the assumption that dark matter is in kinetic equilibrium among itself at all temperatures. From this solution, we are able to derive the constraint $\Lambda_X / a_{12} \sim v_{\rm EW}$, as well as to verify that the dark matter cools to an effectively low temperature by the time of recombination. The temperature of the dark sector as parametrized by $x_{\rm DM}$ is shown in Fig.~\ref{fig:xdm}. One can clearly see from Fig.~\ref{fig:xdm} that the dark matter is hotter than the SM particles after chemical freeze-out era with $x\approx 10^6$. As a result, there is a delayed behavior for the resonant semi-annihilation process to reach the physical pole at $x \sim 10^{11}$ before the recombination era. For $10^{6} < x < 10^{11}$, $x_{\rm DM}$ follows a simple power-law behavior as $x_{\rm DM} \propto x^{2/5}$. After that, the dark matter temperature continues to cool down. At the start of the recombination era with $z=1000$ and $x \approx 6\times 10^{12}$, the corresponding dark matter averaged speed is $v \le 4 \times 10^{-5}$ and well below the resonance pole. 

Maintaining kinetic equilibrium within the dark sector requires the introduction of a new mediator state that interacts with the dark matter. The interaction must be quite large, so the assumption of kinetic equilibrium within the dark matter sector could be difficult to achieve. Then, the dark matter produced in a small time window will begin with a large momentum very close to $3\,M_X/4$. We briefly verify that this does not pose a problem for our assumption that dark matter annihilates at effectively zero velocity around the time of recombination. The semi-annihilation cross section for these states is small at the time they are produced, since they are far from the resonance. They thus redshift as essentially free non-relativistic particles with kinetic energy scaling as $1 / x^2$ so that the hottest dark matter states around the time of recombination have kinetic energy 
\beq
E_{\rm kin} < \frac{9 \, M_X \, x_{\rm kd}^2}{32\,  x_{\rm rec}^2} \sim 10^{-3}~{\rm eV} \ll \delta m \,.
\eeq
Whether or not the dark matter remains in kinetic equilibrium among itself, the cross section at the time of recombination is well approximated by its zero velocity value.

\newpage
\bibliography{AMS2}
\bibliographystyle{JHEP}

 \end{document}